\documentclass[twocolumn]{aastex631}

\shorttitle{Spectro-photometric ICL analysis in RXJ2129}
\shortauthors{Jim\'enez-Teja et al.}

\begin{document}

\title{First joint MUSE, HST, and JWST spectro-photometric analysis of the intracluster light: the case of the relaxed cluster RX J2129.7+0005}

\email{yojite@iaa.es}

\author[0000-0002-6090-2853]{Yolanda Jim\'enez-Teja}
\affiliation{Instituto de Astrof\'isica de Andaluc\'ia--CSIC, Glorieta de la Astronom\'ia s/n, E--18008 Granada, Spain}
\affiliation{Observat\'orio Nacional, Rua General Jos\'e Cristino, 77 - Bairro Imperial de S\~ao Crist\'ov\~ao, Rio de Janeiro, 20921-400, Brazil}

\author[0000-0001-5891-1083]{Antonio Gimenez-Alcazar}
\affiliation{Instituto de Astrof\'isica de Andaluc\'ia--CSIC, Glorieta de la Astronom\'ia s/n, E--18008 Granada, Spain}

\author{Renato A. Dupke}
\affiliation{Observat\'orio Nacional, Rua General Jos\'e Cristino, 77 - Bairro Imperial de S\~ao Crist\'ov\~ao, Rio de Janeiro, 20921-400, Brazil}
\affiliation{Department of Astronomy, University of Michigan, 311 West Hall, 1085 South University Ave., Ann Arbor, MI 48109-1107}
\affiliation{Eureka Scientific, 2452 Delmer St. Suite 100,Oakland, CA 94602, USA}

\author[0009-0004-0063-4414]{Patrick Prado-Santos}
\affiliation{Observat\'orio Nacional, Rua General Jos\'e Cristino, 77 - Bairro Imperial de S\~ao Crist\'ov\~ao, Rio de Janeiro, 20921-400, Brazil}

\author[0000-0001-7299-8373]{Jose M. Vi\'lchez}
\affiliation{Instituto de Astrof\'isica de Andaluc\'ia--CSIC, Glorieta de la Astronom\'ia s/n, E--18008 Granada, Spain}

\author[0000-0002-8742-0643]{N\'icolas O. L. de Oliveira}
\affiliation{Observat\'orio Nacional, Rua General Jos\'e Cristino, 77 - Bairro Imperial de S\~ao Crist\'ov\~ao, Rio de Janeiro, 20921-400, Brazil}

\author[0000-0001-7399-2854]{Paola Dimauro}
\affiliation{INAF – Osservatorio Astronomico di Roma, Via di Frascati 33, 00078 Monte Porzio Catone, Italy}
\affiliation{Observat\'orio Nacional, Rua General Jos\'e Cristino, 77 - Bairro Imperial de S\~ao Crist\'ov\~ao, Rio de Janeiro, 20921-400, Brazil}

\author[0000-0002-6610-2048]{Anton M. Koekemoer}
\affiliation{Space Telescope Science Institute, 3700 San Martin Drive, 
Baltimore, MD 21218, USA}

\author[0000-0003-3142-997X]{Patrick Kelly}
\affiliation{School of Physics and Astronomy, University of Minnesota, 116 Church Street SE, Minneapolis, MN 55455, USA}

\author[0000-0002-4571-2306]{Jens Hjorth}
\affiliation{DARK, Niels Bohr Institute, University of Copenhagen, Jagtvej 155A, 2200 Copenhagen, Denmark}

\author[0000-0003-1060-0723]{Wenlei Chen}
\affiliation{Department of Physics, Oklahoma State University, 145 Physical Sciences Bldg, Stillwater, OK 74078, USA}

\begin{abstract}
We present the most detailed spectrum of the intracluster light (ICL) in an individual cluster to date, the relaxed system RX J2129.7+0005, at $z\sim 0.234$. Using 15 broad-band, deep images observed with HST and JWST in the optical and the infrared, plus deep integral field spectroscopy from MUSE, we computed a total of 3696 ICL maps spanning the spectral range $\sim 0.4-5$ $\mu$m with our algorithm CICLE, a method that is extremely well suited to analyzing large samples of data in a fully automated way. We used both parametric and non-parametric approaches to fit the spectral energy distribution of the ICL and infer its physical properties, yielding a stellar mass $log_{10}(M_*/M_{\odot})$ between $11.5-12.7$ and an average age between $9.7-10.5$ Gyr, from CIGALE and Prospector results. This implies that the ICL in RX J2129.7+0005 is, on average, older than that of disturbed clusters, suggesting that the contribution from different stellar populations to the ICL are at play depending on the cluster's dynamical state. Coupled with X-ray observations of the hot gas distribution, we confirm the relaxed state of RX J2129.7+0005, showing clear signs of sloshing after a last major merger with a high-mass ratio satellite that could have happened $\sim 6.6$ Gyr ago in a relatively radial orbit. The presence of substructure in the ICL, such as shells, clouds with different densities and a certain degree of boxyness, and a clump, supports this scenario.
\end{abstract}


\section{Introduction} \label{sect:intro}

The intracluster light (ICL) is for certain the least studied luminous component of the galaxy clusters. The ICL is defined as the light emitted by those stars that are gravitationally bound by the cluster potential but do not belong to any galaxy in particular. These stars are primarily freed from their progenitor host galaxies by the numerous interactions occurred in the cluster during its assembly history \citep[e.g.][]{demaio2018}, although they can also be born in situ \citep{puchwein2010,sun2010,barfety2022}. As privileged witness of all these interactions, the ICL carries the imprint of the past and present dynamical history of the cluster. Observers usually trace back this history by analyzing properties like the radial color of the ICL, its morphology, or the ICL fractions \citep[e.g., ][]{morishita2017,Jimenez-Teja2018,jimenez-teja2024}. The best relic of past and ongoing mergers, though, is the presence of substructure in the ICL, like clumps, tidal features, shells, and others. However, the low surface brightness of the ICL makes the observation of these features extremely complex and just limited to nearby clusters, with very few exceptions \citep[e.g., ][]{adami2013}. \\

The advent of James Webb Space Telescope (JWST), with extraordinary sensitivity and spatial resolution in the infrared (IR), has revolutionized our picture of the ICL. The most clear example is the galaxy cluster SMACS J0723.3-7327, at $z = 0.39$, the first cluster observed by JWST as part of its Early Release Observations. Although it was a well-known system, widely studied as one of the targets of the Reionization Lensing Cluster Survey \citep[RELICS, ][]{coe2019}, the superb-quality JWST images revealed a wealth of substructures that had remained unnoticed by HST. JWST unveiled a myriad of compact sources and globular clusters tracing the ICL, as well as signs of substructure like tidal streams, plumes, and loops \citep{lee2022,faisst2022,montes2022,diego2023}. This picture suggests that the ICL distribution is substantially more complex than we thought, especially in the case of dynamically active (merging) clusters as SMACS J0723.3-7327.\\

With the launch of JWST, the ICL has, unexpectedly, attracted a lot of attention. In the hunt for high-redshift sources - frequently lensed by foreground galaxy clusters and, in many cases, intrinsically faint - JWST photometry can be significantly impacted by the glow of the ICL. Recent works developed special techniques to model not only the background but also the ICL locally, in small regions that surround compact, lensed galaxies and star clusters \citep[e.g.][]{mowla2022,welch2022,hsiao2023,meena2023,adamo2024}. As a consequence, this new era in the ICL studies opens the door to a more detailed, and more comprehensive, examination of the ICL: the search for substructure that can widen our comprehension of the hierarchical process of formation and evolution of the clusters and their internal dynamics. \\

\begin{figure*}
\includegraphics[width=0.511\textwidth]{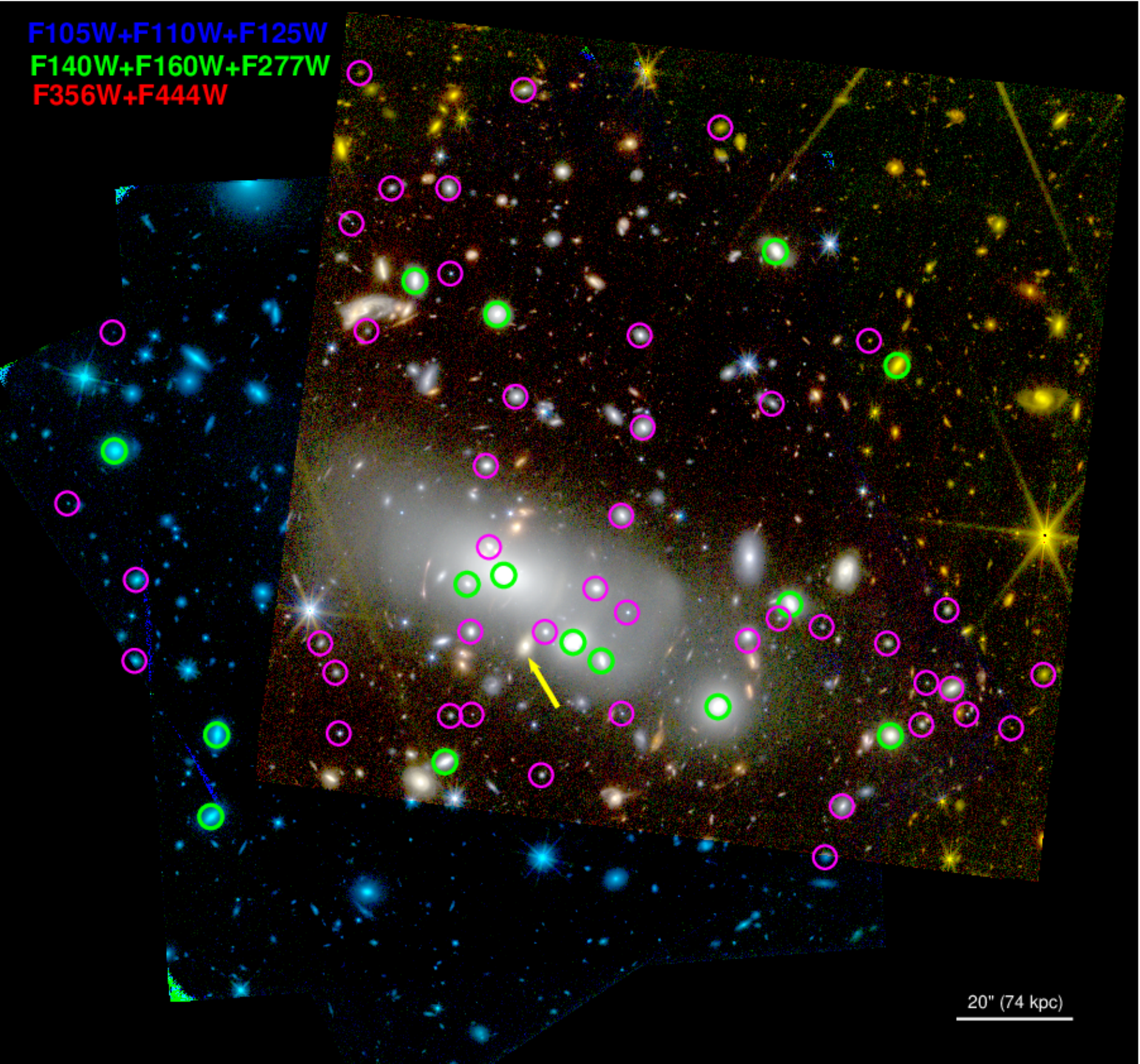}
\includegraphics[width=0.489\textwidth]{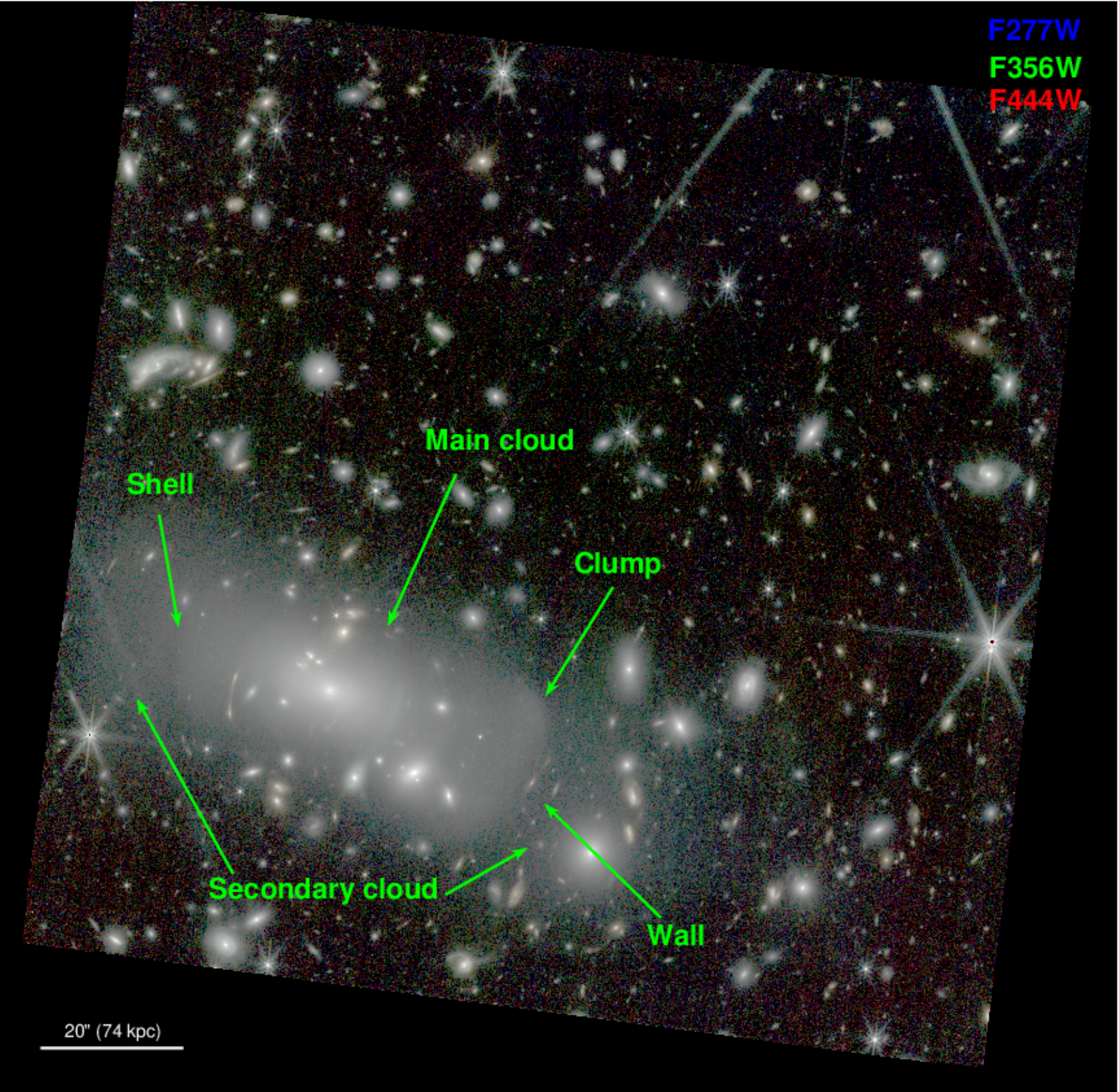}
\caption{Colored images of RXJ2129 made with Trilogy \citep{coe2012}  and different IR filters. Left: false color image generated from HST/IR and JWST/LW filters. Circles indicate cluster galaxy members that are spectroscopically (green) and photometrically (magenta) identified. The yellow arrow points to a background galaxy at $z\sim 0.6$, which has a reddish tail that overlaps the ICL in projection . Right: false color image generated from the JWST/LW channel filters solely. The different color rendering highlights fainter objects and the ICL, which spreads far beyond the BCG. We identify a denser cloud of ICL that overlaps with the BCG in projection (main cloud) and with several shells. A fainter ICL extension spreads out both west and east to the BCG (secondary cloud). Several substructures, like a clump, are clearly visible by eye. North is up, east is left, {as in all the following figures.}} \label{fig:RGB_original}
\end{figure*}

In this work, we conduct, for the first time, a spectro-photometric analysis of the entire ICL in a cluster. To date, only a handful of clusters had their ICL studied through spectroscopy and, thus, their stellar properties reliably characterized in detail \citep{coccato2011,toledo2011,edwards2016,melnick2012,adami2013}. We use a combination of MUSE, HST, and JWST data of the galaxy cluster RX J2129.7+0005, at $z=0.234$, a relaxed system for which an intriguing ICL distribution can be observed by eye from a simple color composite image (see Fig. \ref{fig:RGB_original} right). The main ``cloud'' of ICL appears as an extended envelope that surrounds the brightest cluster galaxy (BCG) and some nearby, smaller galaxies. It is rather symmetrical and aligned with the BCG, elongated with a main axis with position angle of $\sim 70^{\circ}$. This main cloud spreads into a less dense ICL extension, which we will call secondary cloud. Interesting features, like shells, a clump, or a certain degree of boxyness are observed in the ICL.\\


This paper is organized as follows. Section \ref{sect:data} describes the optical and infrared imaging and integral field spectral data used in the analysis of the ICL, along with the Chandra observations used to study the X-ray counterpart of the hot gas. The processing of the different data to generate the ICL maps is described in Sect. \ref{sect:ICL_maps} for HST and JWST, and in Sect. \ref{sect:ICLmaps_MUSE} for MUSE. These maps are used to {derive} the physical properties of the ICL in Sect. \ref{sect:SEDfitting}. The X-ray analysis is fully described in Sect. \ref{sect:x-rays}. Finally, {our} results are discussed in Sect. \ref{sect:discussion} and the main conclusions summarized in Sect. \ref{sect:conclusions}. Throughout this paper, we will assume a standard $\Lambda$CDM cosmology with $H_0=70$ km s$^{-1}$ Mpc$^{-1}$, $\Omega_m=0.3$, and $\Omega_{\Lambda}=0.7$. All the magnitudes are referred to the AB system.\\

\section{Data} \label{sect:data}
The analysis of the stellar properties of the ICL is derived from HST and JWST optical and infrared imaging, along with integral field spectra from MUSE. We combine these data with X-ray observations to draw a complete picture of the dynamics of RX J2129.7+0005.  {All the HST and JWST data used in this paper can be found in the Mikulski Archive for Space Telescopes (MAST) at the Space Telescope Science Institute. The specific observations analyzed can be accessed via \dataset[doi: 10.17909/tpsv-zp97]{https://doi.org/10.17909/tpsv-zp97}.}  \\

\subsection{HST and JWST data} \label{sect:data:HST_JWST}
RX J2129.7+0005 (R.A.: 21$^{\rm h}$29$^{\rm m}$40$^{\rm s}$, Dec.:+00$^{\circ}$05'18.8'', $z=0.234$; hereafter, RXJ2129) was one of the 25 massive clusters observed by the Cluster Lensing And Supernova survey with Hubble \citep[CLASH,][]{postman2012}, between 2010 and 2013. Its potentially relaxed state, given the symmetry of its hot gas distribution, was among the selection criteria to consider RXJ2129 as one of these 25 clusters. The CLASH collaboration observed RXJ2129 with the Hubble Space Telescope (HST) in four ultraviolet and five infrared (IR) bands with the WFC3 (UVIS and IR channels, respectively), and in eight optical bands with the ACS (see Table \ref{table:observations}). We downloaded the publicly available mosaics from the CLASH website\footnote{https://archive.stsci.edu/prepds/clash/}. These mosaics were produced with MultiDrizzle \citep{koekemoer2002,koekemoer2011}, from the calibrated files. Images were corrected for bias stripping effects, charge transfer efficiency degradation, geometric distortion, cosmic rays, and bad pixels. In \cite{jimenez-teja2021}, we proved that the standard flat-field correction derived from the HST pipeline, although not optimal for low-surface-brightness studies, does not have a significant impact on the photometry of the ICL. For the case of RXJ2129 and given that our primary goal is to unveil the fine ICL substructure and details, we used the mosaics with the optimal resolution for the ACS and WFC3/UVIS data, which is 0.03 arcsec/pixel. We did not use the WFC3/UVIS images because  the ICL was not detected in them. We also rejected the F555W ACS mosaic because the BCG and, therefore, the ICL, lie outside the field of view. In Table \ref{table:observations}, we list the 12 HST filters used, along with the exposure times. \\

\begin{table*}
\centering
\begin{tabular}{lcccc}
Telescope/Instrument/Channel & Filter & Exposure Time & Pixel scale & ICL mag \\
 & & [s] & [arcsec/pixel] & [mag] \\
   \hline
HST/ACS & F435W & 3910 & 0.03 & $19.25\pm 1.19$\\
HST/ACS & F475W & 3728 & 0.03 & $18.72\pm 0.79$\\
HST/ACS & F606W & 3870 & 0.03 & $17.69\pm 0.40$\\
HST/ACS & F625W & 3728 & 0.03 & $17.43\pm 0.48$ \\
HST/ACS & F775W & 7792 & 0.03 & $16.56\pm 0.25$ \\
HST/ACS & F814W & 7866 & 0.03 & $16.79\pm 0.24$ \\
HST/ACS & F850LP & 15084 & 0.03 & $16.82\pm 0.29$\\
HST/WFC3/IR & F105W & 2614 & 0.03 & $16.26\pm 0.30$ \\ 
HST/WFC3/IR & F110W & 2414 & 0.03 & $16.14\pm 0.24$ \\ 
HST/WFC3/IR & F125W & 3420 & 0.03 & $15.97\pm 0.24$ \\
HST/WFC3/IR & F140W & 2311 & 0.03 & $15.96\pm 0.27$ \\
HST/WFC3/IR & F160W & 6238 & 0.03 & $15.77\pm 0.19$\\
JWST/NIRCam/SW & F115W & 8245 & 0.02 & ... \\
JWST/NIRCam/SW & F150W & 19927 & 0.02 & ...\\
JWST/NIRCam/SW & F200W & 8245 & 0.02 & ... \\
JWST/NIRCam/LW & F277W & 2061 & 0.04 & $15.70\pm 0.08$ \\
JWST/NIRCam/LW & F356W & 4981 & 0.04 & $16.06\pm 0.06$\\
JWST/NIRCam/LW & F444W & 2061 & 0.04 & $16.33\pm 0.09$\\
\hline
\end{tabular}
\caption{HST and JWST optical and IR observations used for this work, along with the corresponding exposure times. The {fourth} column indicates the photometry of the ICL for total region described in {Sect. \ref{sect:SEDfitting}} and outlined in Fig. \ref{fig:regions}. We lack photometric measurements for the three SW channel images because some wisps contaminated the northwestern side of the ICL.} \label{table:observations}
\end{table*}

\begin{figure}
\includegraphics[width=0.49\textwidth]{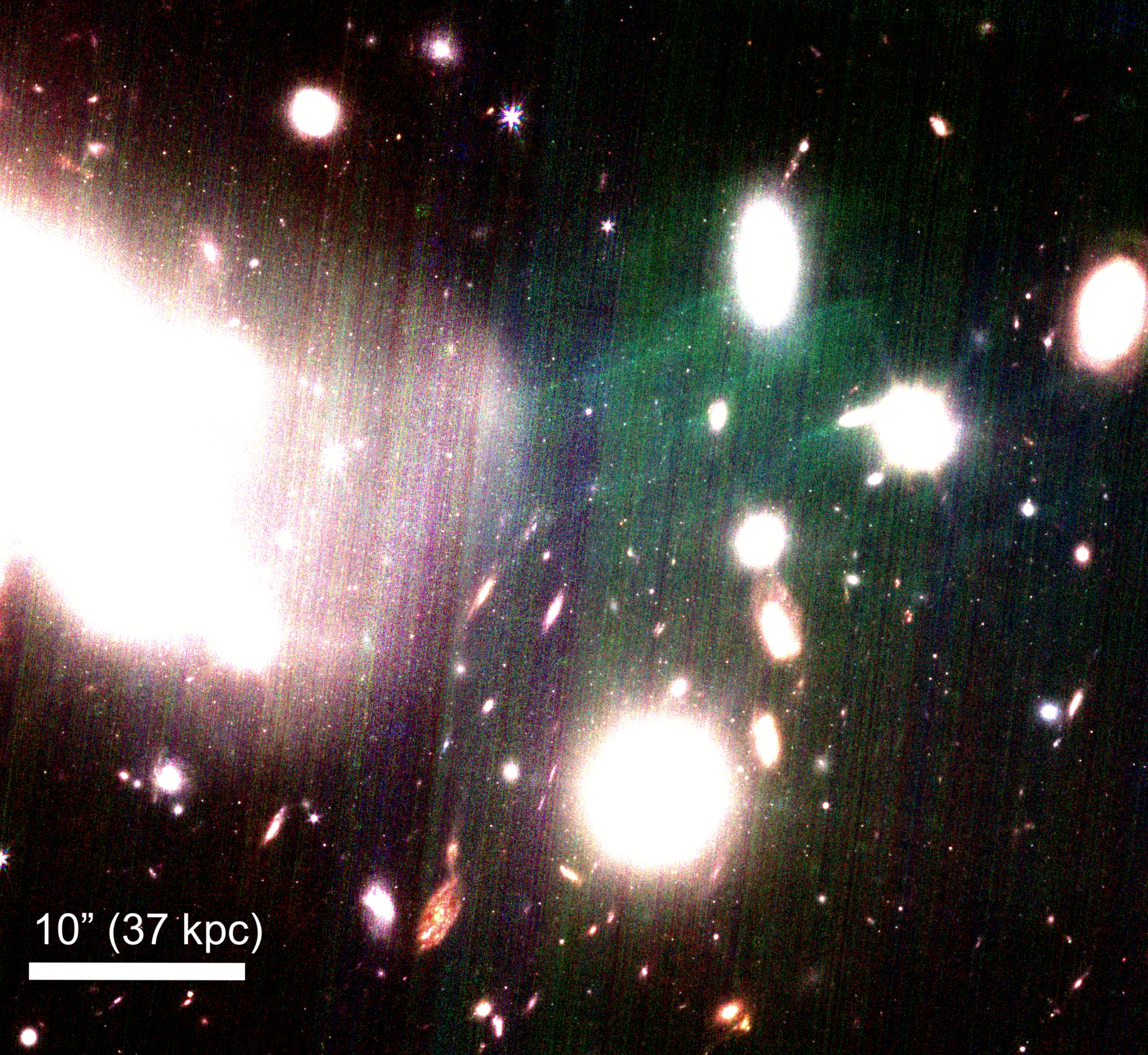}
\caption{Colored images of RXJ2129 made with Trilogy and the F150W, F200W, and F277W JWST filters, where some wisps are clearly visible, featured as a green filamentous structure west to the BCG.} \label{fig:wisps}
\end{figure}

RXJ2129 was also one of the galaxy clusters targeted by JWST, as part of the Director's Discretionary Time (DD-2767, PI: Kelly). It was observed with the NIRCam in three short- (SW) and three long-wavelength-channel (LW) filters, also listed in Table \ref{table:observations}. We identified some wisps west to the BCG, a {well-known} artifact that occasionally affects images observed with NIRCam SW channel filters (see Fig. \ref{fig:wisps}). Wisps are diffuse stray light features, which look like faint filamentous structures. They have low surface brightness and are variable in shape and position, as a function of telescope pointing. This makes the use of templates or other more sophisticated techniques to remove them \citep{robotham2023} not completely efficient when wisps are located in a region with ICL and with limited other exposures to construct reliable wisp templates. For this reason, we do not use the SW-channel filters in our photometric analysis. \\

The JWST NIRCam images were pre-processed with a custom calibration that is fully described in \cite{williams2023}. The individual exposures were firstly processed with the standard CALWEBB software \citep{bushouse2019} and then astrometrically aligned using a catalog constructed from previous images observed with SuprimeCam at Subaru Telescope by the Cluster Lensing and Supernova Survey with Hubble program \citep[CLASH, ][]{postman2012}. Further steps included reference pixel correction, flat-fielding, and bias corrections, and background subtraction to the individual exposures. These were coadded and astrometrically aligned into the final mosaics at a pixel scale of 0.02 arcsec/pixel for the SW-channel images, and 0.04 arcsec/pixel for the LW-channel images. \\
 
False color images of the IR emission of RXJ2129 are shown in Fig. \ref{fig:RGB_original}, built using Trilogy {\citep{coe2012}} and two different combinations of JWST filters. {Trilogy is a code that creates color images by applying a logarithmic scale to the red, green, and blue channels independently, with weights that are optimized to enhance faint features without saturating bright structures. The scalings are mainly dependent on two input parameters: the percentage of pixels that are allowed to saturate and the output luminosity of the noise. In order to make the ICL more visible, we usually choose very high levels of saturation ($\geq 10$\%).} The superb sensitivity and depth of the JWST data allow us to see the ICL of RXJ2129 directly by eye, previous to processing the images. The ICL is clearly concentrated around the BCG, in a dense, elongated cloud that we call main cloud. At first inspection, this cloud shows an overall symmetric distribution, although we appreciate some remarkable differences comparing its eastern and western sides. Towards the east of the BCG, the ICL diffuses slowly and progressively into the intracluster space, while to the west of the BCG, the ICL shows a significant boxyness and a clear ``wall'' or front. {We remark} the presence of a small region in the north end of this front where the ICL appears clearly denser, which we call clump from now on. It is worth mentioning the presence of some other substructures in the main cloud as shells and the tail of a galaxy southwest to the BCG (visible in Fig. \ref{fig:RGB_original} left). The redshift calculated with the HST photometry \citep{molino2017} places this galaxy in the {background}, at $z\sim 0.6$, thus its reddish tail is not the result of a possible interaction with the cluster potential nor it contributes to the ICL budget. \\

\subsection{MUSE data} \label{sect:data:MUSE}

In addition to the imaging, RXJ2129 was also observed with the Multi Unit Spectroscopic Explorer (MUSE), at the VLT UT4 telescope. MUSE creates a datacube of 3681 images that sample almost the full optical domain ($\sim 475 - 935$ nm) with a mean spectral resolution of 3000. Coupled with spatial resolution 0.2 arcsec/pixel, MUSE is an excellent instrument to analyze the ICL via a spectro-photometric approach. RXJ2120 belongs to the MUSE-DEEP collection, a set of targets that have been intensively observed by MUSE and for which extraordinary deep data cubes are available, combining all existing observations. In particular, the MUSE-DEEP datacube of RXJ2129 has a total integration time of all exposures of 20,720 sec.\\

\begin{figure}
\centering
\includegraphics[width=0.5\textwidth]{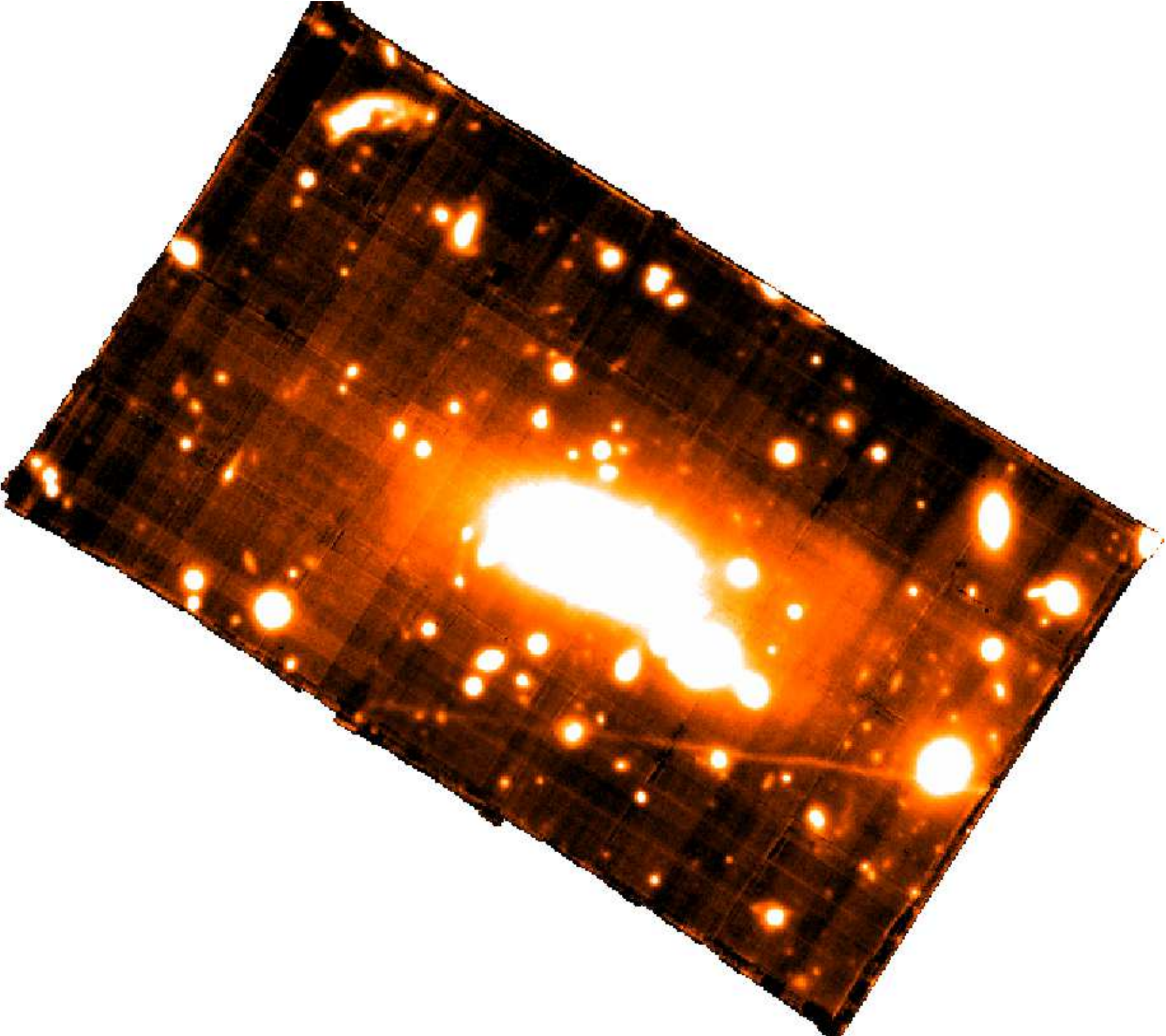}
\caption{MUSE white image of RXJ2129} \label{fig:MUSEimage}
\end{figure}

All MUSE-DEEP datacubes provided are fully reduced by the Quality Control Group at ESO. Steps of the reduction pipeline include sky subtraction and alignment and combination of all visits made by different programmes\footnote{http://www.eso.org/rm/api/v1/public/releaseDescriptions/102}. Fig. \ref{fig:MUSEimage} shows a white image of RXJ2129, created by median-averaging all frames of the reduced datacube in the spectral axis.  \\

\subsection{Chandra data} \label{sect:data:Xray}

All archival data from Chandra available for RXJ2129 were used to study the distribution of the hot gas in X-rays. There are two separate observations of RXJ2129 (OBS ID: 0552 and 9370) taken with the Advanced CCD Imaging Spectrometer (ACIS-I), in October 2000 (PI: Van Speybroeck) and April 2009 (PI: Allen), respectively. We used CIAO 4.14 to run the standard recommended data processing steps, reprocessing all the data with the script \textit{chandra\_repro}, which creates new bad pixel files and new level 2 event files. No flare periods were detected and the total effective exposure is 40.1 ksec. We created a merged event file, reprojecting all observations to a common tangent point using the \textit{reproject\_obs} tool. This was used solely for dealing with point source removal and define the best extraction regions. For the image analysis, we used the deeper exposure pointing, avoiding uncertainties related to a slight offset between aimpoint configurations (\textit{CIAO} guidelines\footnote{ cxc.cfa.harvard.edu/ciao/ahelp/reproject\_obs.html}). \\

\section{Generation of the ICL maps: HST and JWST} \label{sect:ICL_maps}

The presence of galactic cirri, diffuse dust structures within our own Milky Way, can complicate the identification and characterization of the ICL. Galactic cirri are interstellar dust clouds, predominantly observed at high galactic latitudes $({\it{b}} \geq 20^\circ)$. As RXJ2129 is far from the plane of our galaxy (Galactic coordinates: $l=053.660762^{\circ}$, $b=-34.476917^{\circ}$) and it has relatively low galactic reddening of $E(B-V)=0.035$ (from https://irsa.ipac.caltech.edu/applications/DUST/), our images have a low probability of being contaminated by our galaxy. We also checked whether the filaments have a far infrared counterpart, since it is where galactic cirri typically have their emission peak due to their low temperature \citep{low1984,appleton1993}. We downloaded data from surveys such as the Wide-field Infrared Survey Explorer (WISE), the Infrared Astronomical Satellite (IRAS), and Planck \citep{neugebauer84,wright10,planck2020}. A comparative analysis of the full-sky maps provided by those surveys revealed the absence of any significant projected overlap between the ICL and the galactic cirri. \\


Once we ruled out possible contamination from galactic cirri, we generated the ICL maps using the CHEFs Intracluster Light Estimator \citep[CICLE, ][]{Jimenez-Teja2016,Jimenez-Teja2018}. CICLE models and removes the luminous distribution of all galaxies in the image in two dimensions using orthonormal mathematical bases called CHEFs, built in polar coordinates using Chebyshev rational functions (in the radial coordinate) and Fourier series (in the azimuthal coordinate) \citep{Jimenez-Teja2012}. By definition of mathematical basis, CHEFs are capable of fitting any surface distribution that lies within its domain of definition, being this formed by all functions that are smooth. This means that CHEFs cannot fit sharp edges as those arisen from saturated pixels, diffraction spikes of stars, or galaxies that are cut near the borders of the image. Thus, the first steps of CICLE are masking out these sharp objects and detecting all the unmasked sources with SExtractor \citep{Bertin2002,Bertin2012}. CICLE then builds a model for each one of these sources with CHEFs. By construction, CHEFs are very flexible and able to fit any smooth two dimensional distribution with a low number of components (typically, a maximum number of 10 Chebyshev rational functions and 10 Fourier modes). However, for the particular case of the JWST images, given their superb resolution and small pixel scale (0.02 arcsec/pixel), we allowed the CHEF models to be constructed with a maximum number of 15 components for each polar coordinate, to better recover the finest details. \\

The CICLE code determines the extension of each CHEF model, being delimited by the curve of points where the stellar halo of the fitted galaxy either submerges into the sky or converges asymptotically to a certain limit. These are simple criteria to implement for satellite galaxies, considering as satellite any galaxy in the region of the cluster (member or not), which is not the central BCG. However, given the spatial coincidence in projection of the BCG and the ICL and the extended envelope that usually surrounds this special galaxy, it is not straightforward to calculate the transition from the BCG- to the ICL-dominated region. CICLE incorporates a special recipe to find these limits, which are defined as those points where the curvature (intuitively understood as the change in the slope) of the BCG+ICL system is highest. It is important to note that this transition is calculated in two dimensions, thus avoiding any intrinsic assumption of symmetry, neither for the BCG nor the ICL. In \citet{Jimenez-Teja2016}, we estimated the accuracy of this separation and its impact on the final ICL fractions measured, by testing CICLE against simulations. We found an excellent performance with an error of 1\% for the average clusters at $z\sim 0.25$, close to the $z\sim 0.23$ RXJ2129 cluster analyzed here. More recently, CICLE participated in a blind challenge to cross-compare all current algorithms of ICL analysis. The challenge was run over a set of mock clusters, generated from four different suites of cosmological hydrodynamical simulations, with the observational characteristics expected for the Vera C. Rubin Observatory's Legacy Survey of Space and Time data after ten years of operations \citep{brough2024}. CICLE's performance surpassed that of all other techniques, with the closest measurements to those of the simulators (the most precise), the smallest scatter (the most reliable), and among the least affected by {projection} effects (robustness). Moreover, CICLE is one of the few techniques fully automated and, therefore, prepared to process large samples of clusters, which is key to study the ICL using datacubes from integral field units (IFUs, see Sect. \ref{sect:data:MUSE}). \\

\begin{figure*}
\centering
\includegraphics[width=.48\textwidth]{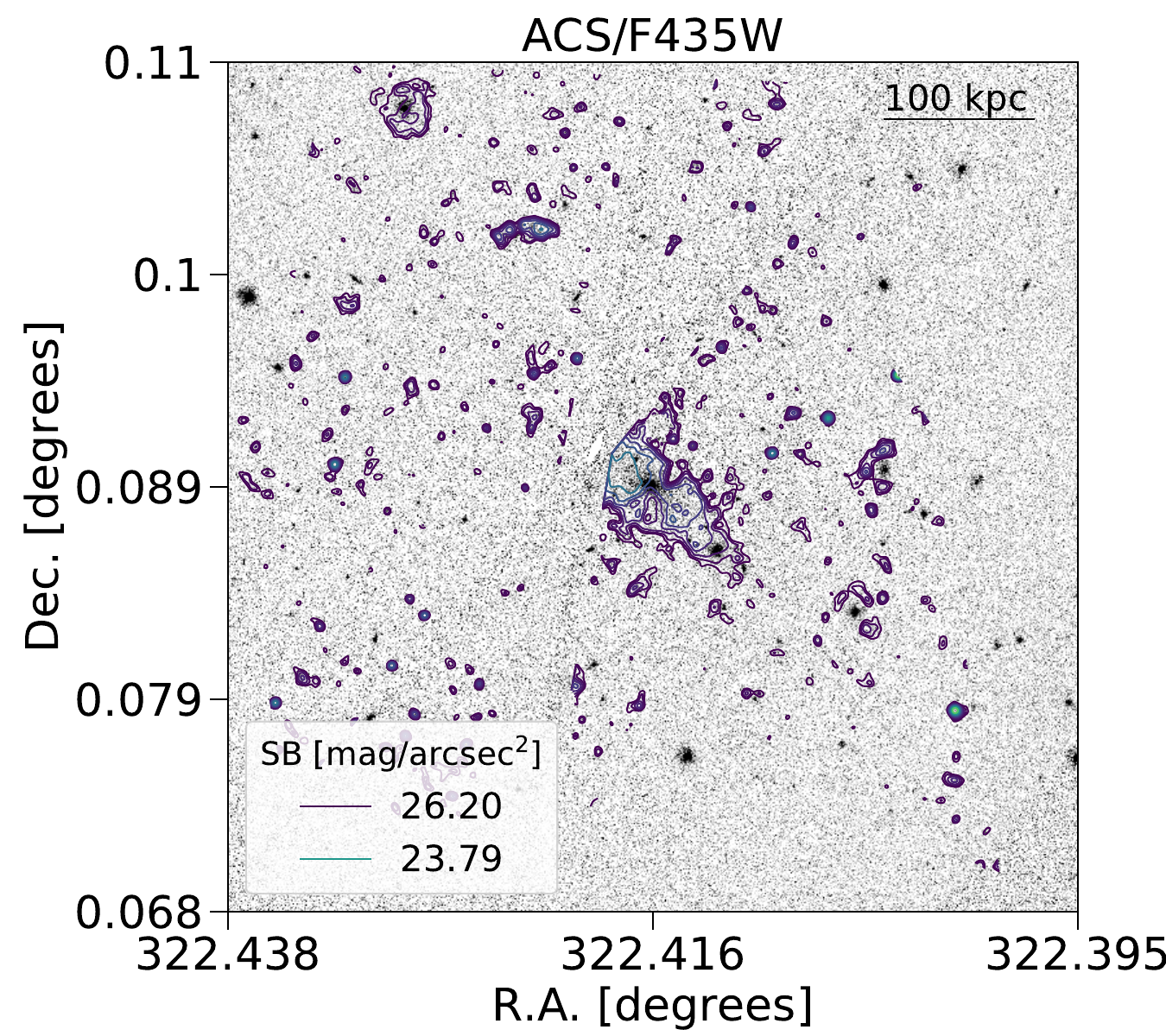}\hfill\includegraphics[width=.48\textwidth]{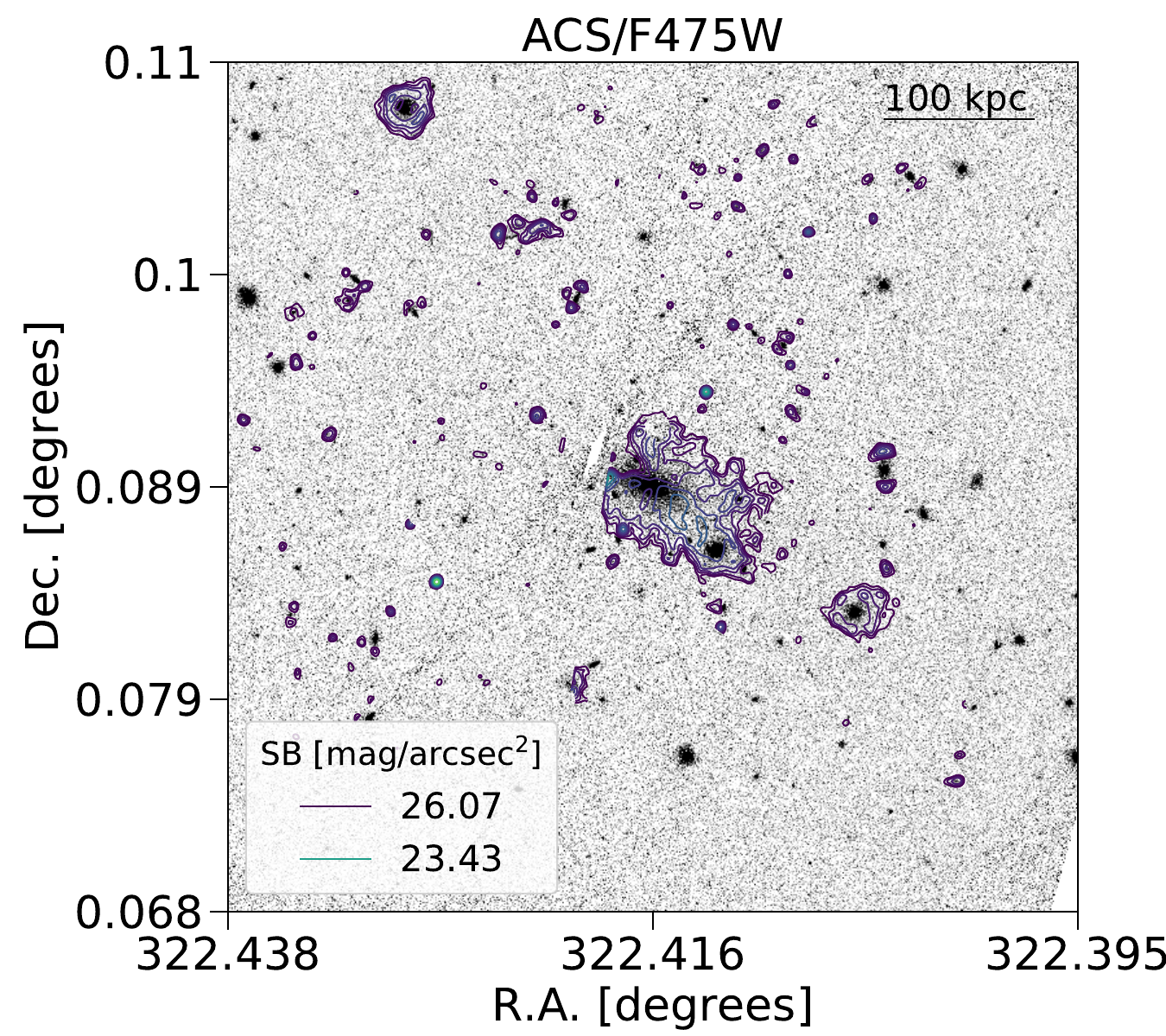}\hfill \\
\includegraphics[width=.48\textwidth]{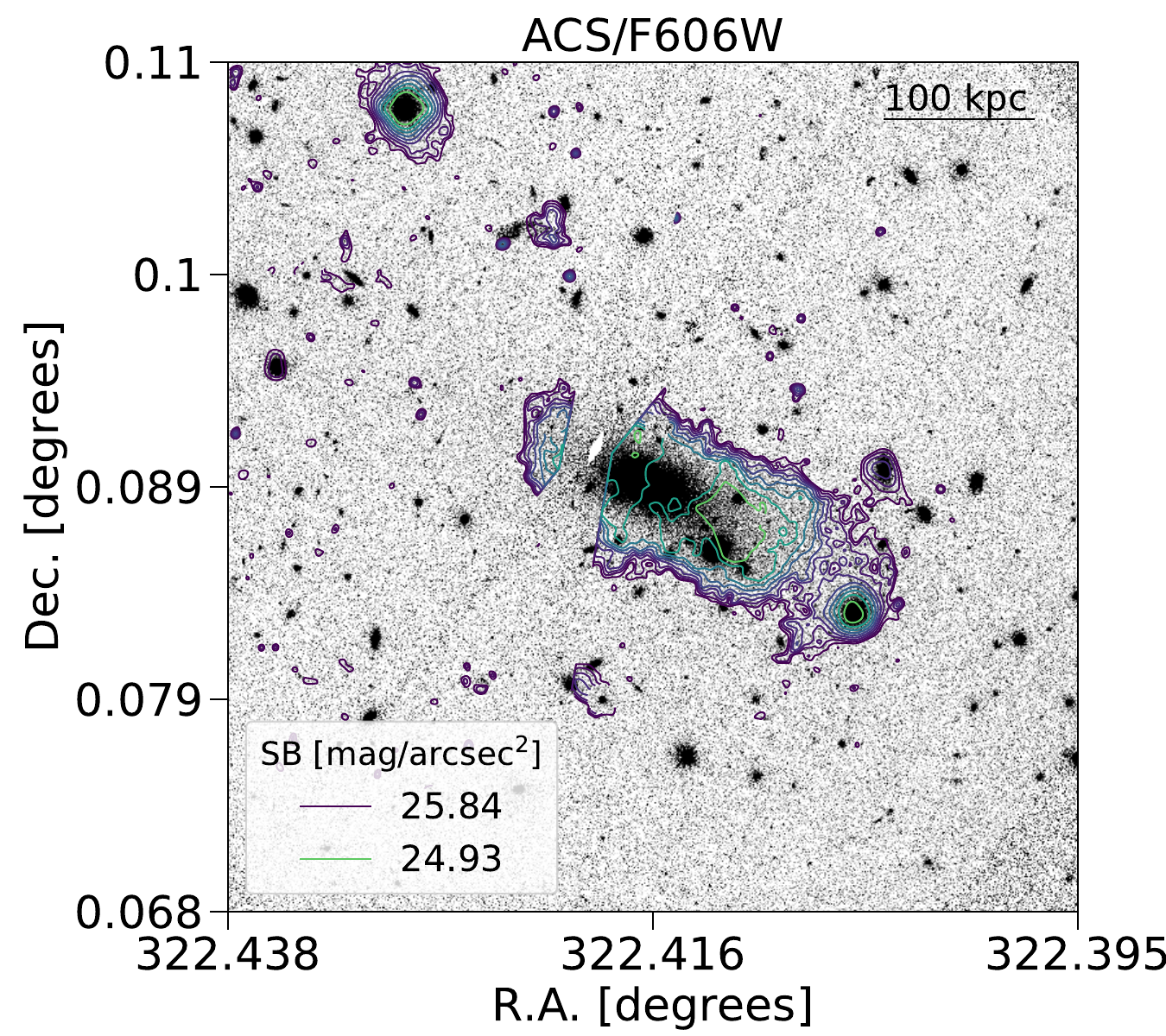}\hfill\includegraphics[width=.48\textwidth]{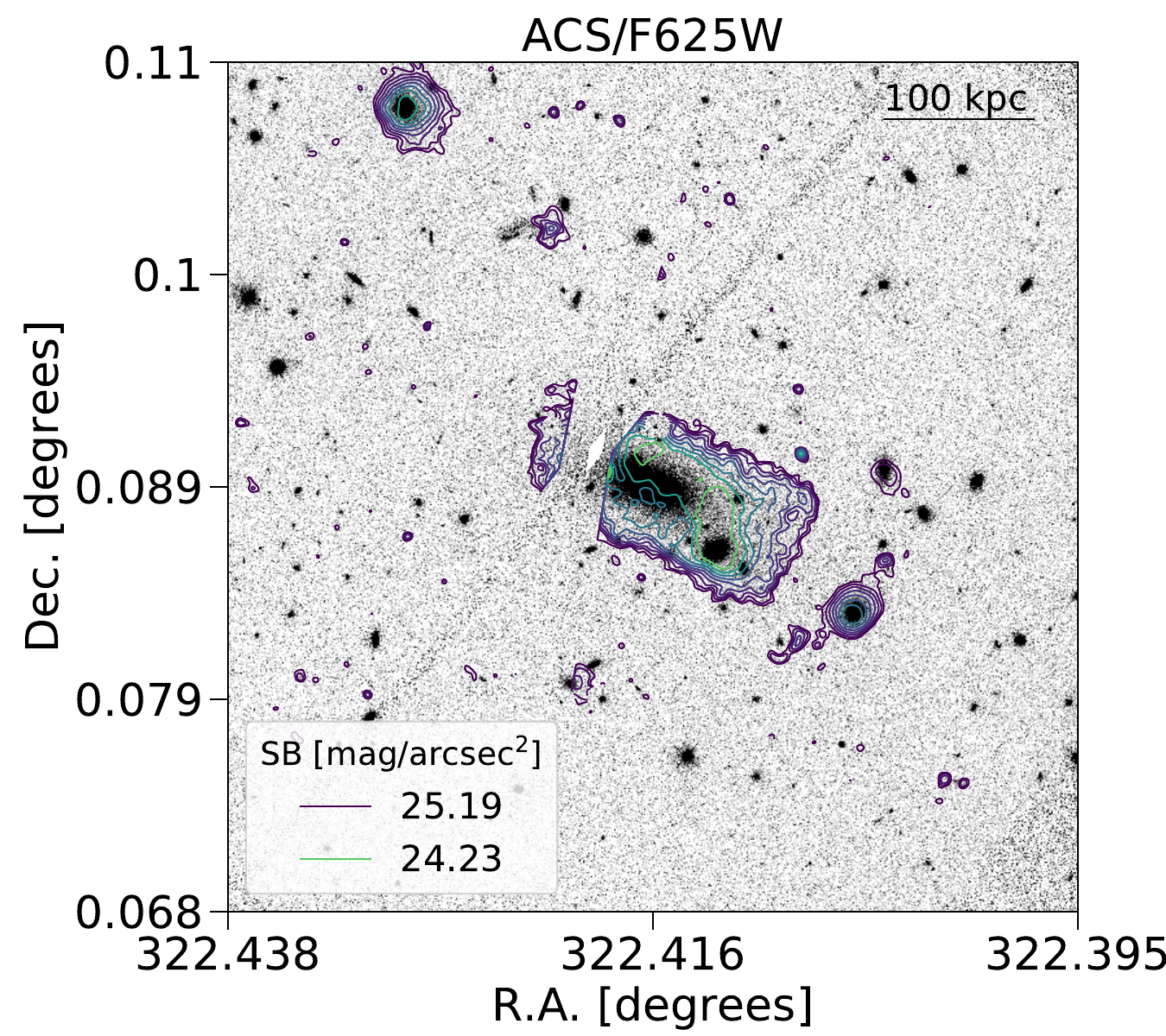}\hfill \\
\includegraphics[width=.48\textwidth]{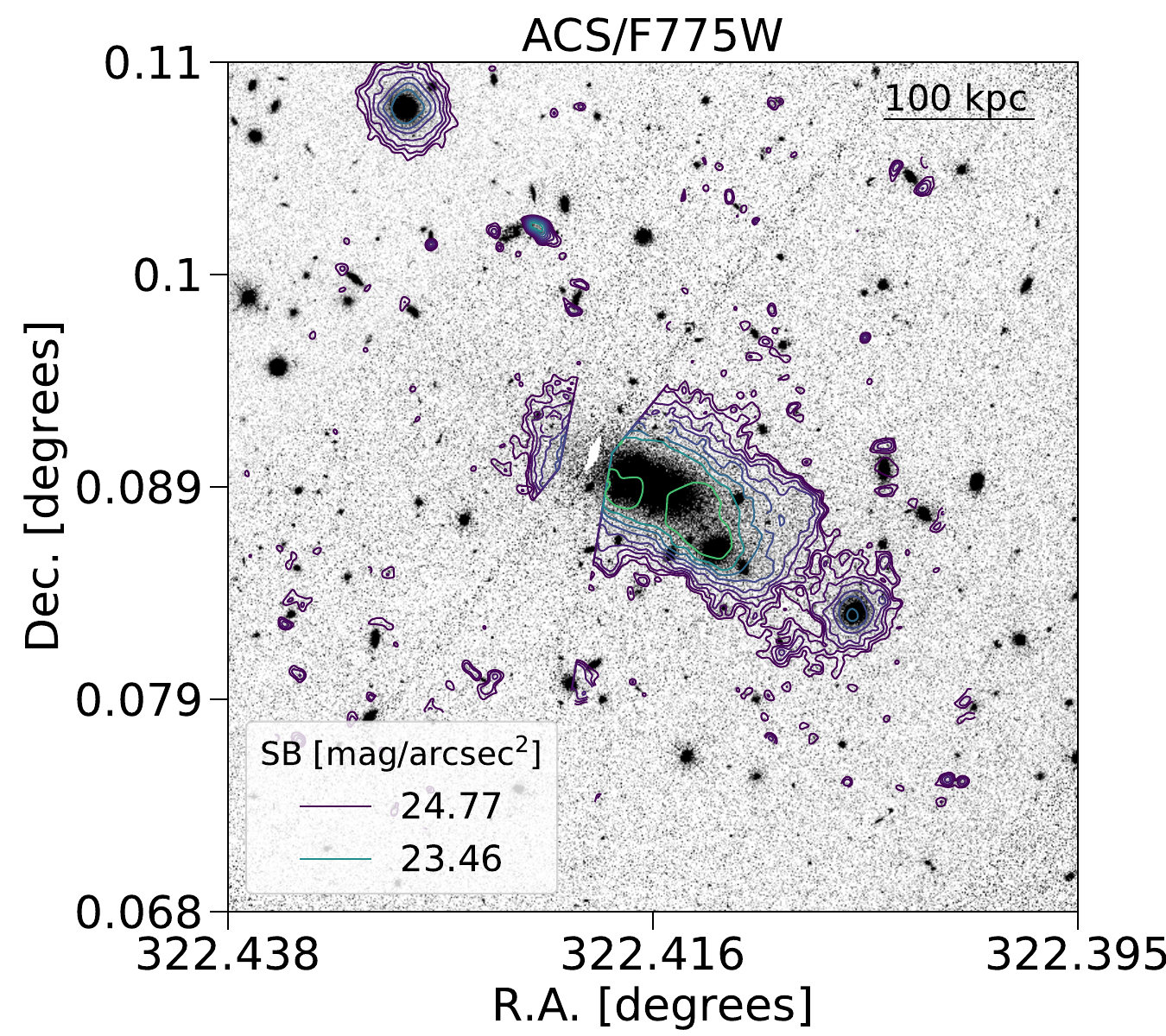}\hfill\includegraphics[width=.48\textwidth]{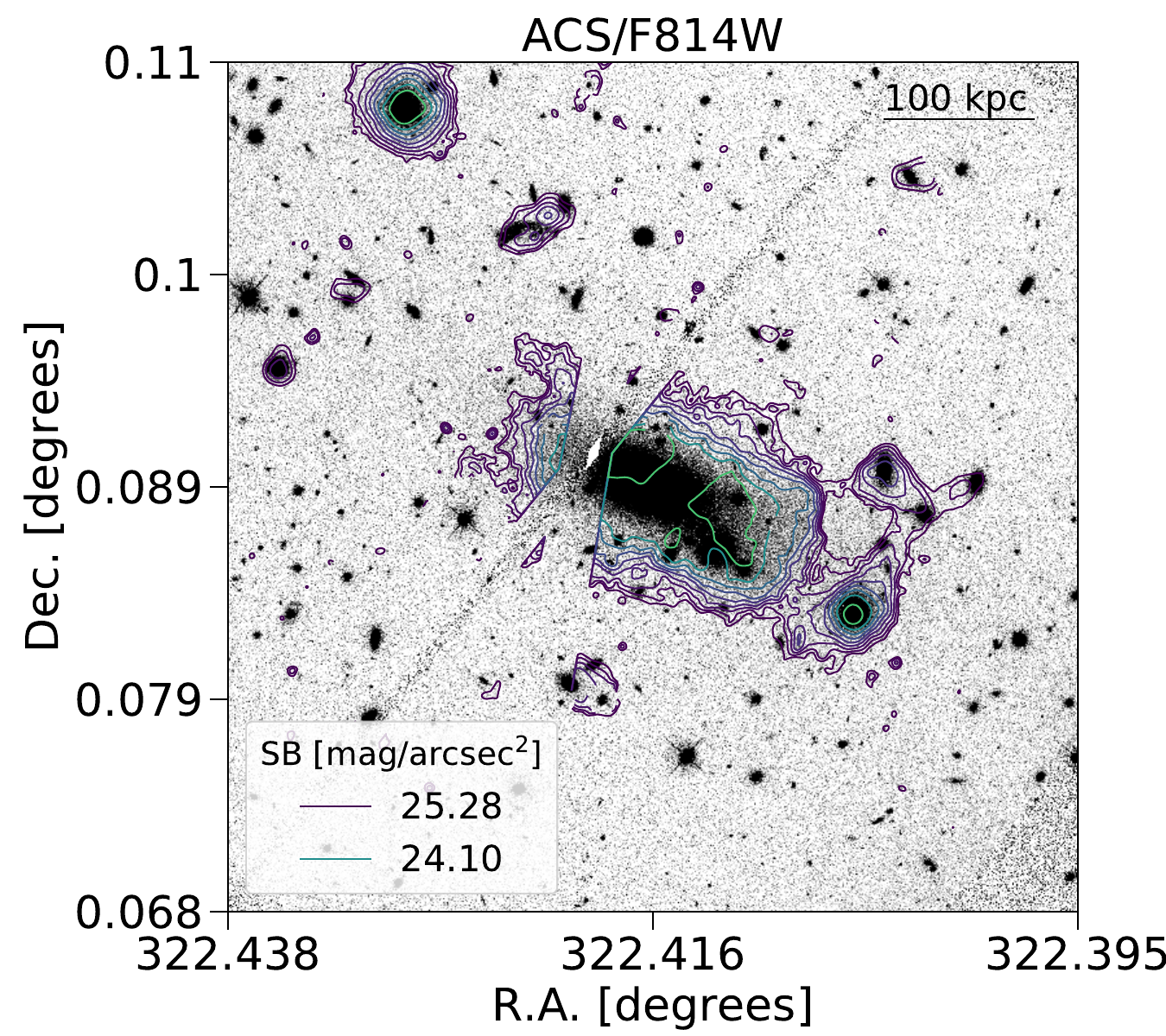}\hfill \\
\caption{ICL isocontours superimposed over the original images in the twelve HST filters considered for this work. } \label{fig:ICL_contours_1}
\end{figure*}

\begin{figure*}
\centering
\includegraphics[width=.48\textwidth]{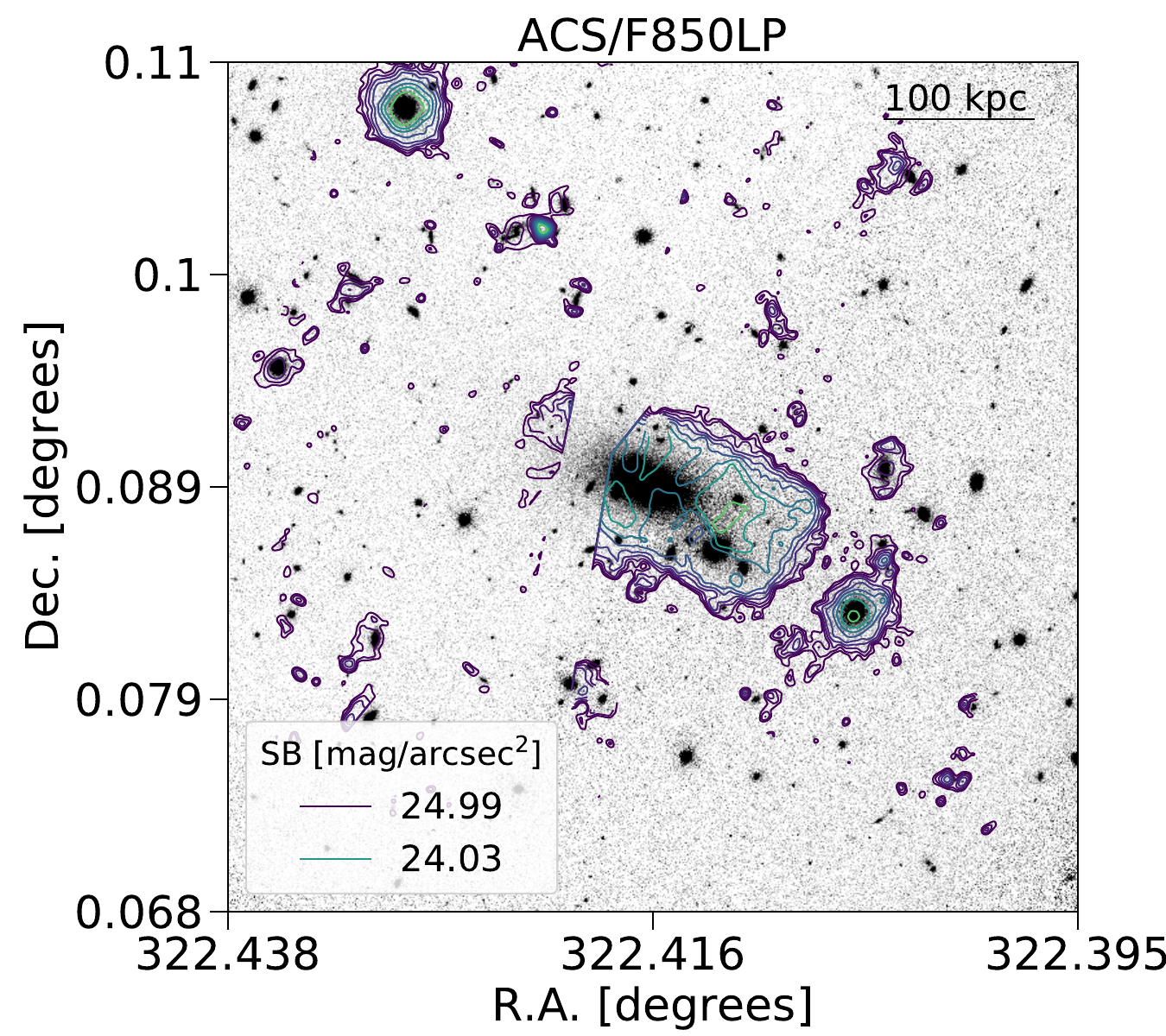}\hfill\includegraphics[width=.48\textwidth]{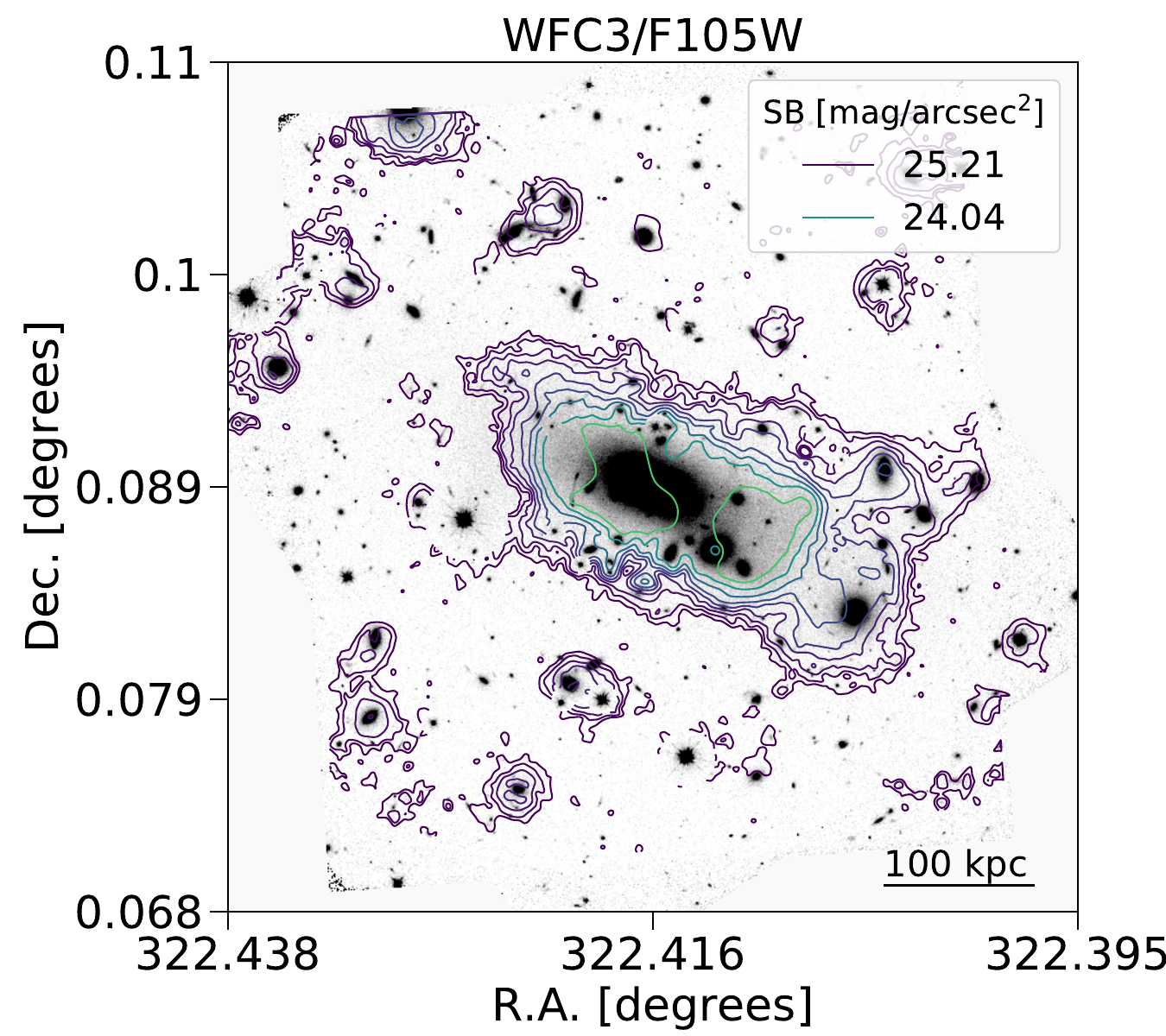}\hfill \\
\includegraphics[width=.48\textwidth]{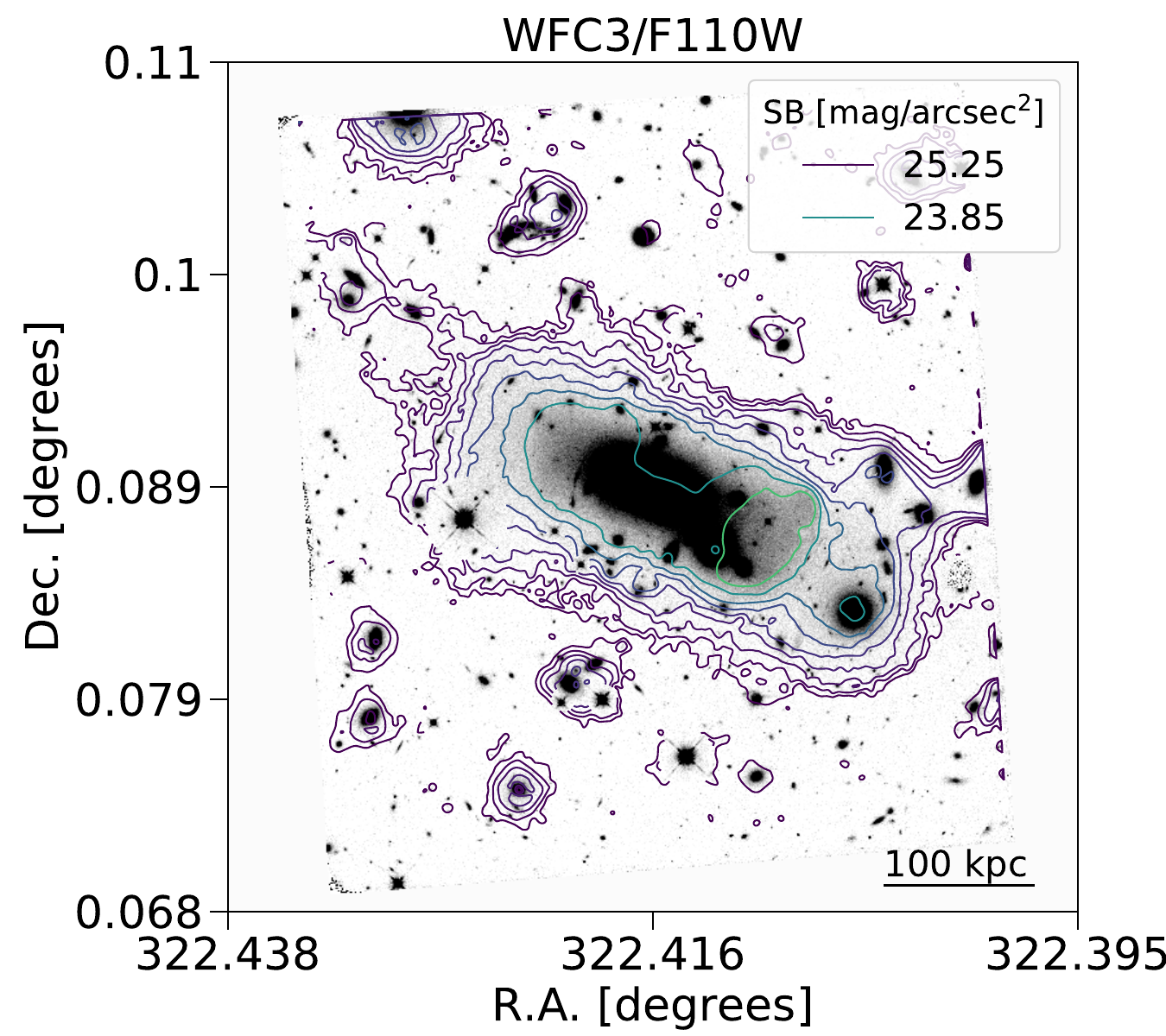}\hfill\includegraphics[width=.48\textwidth]{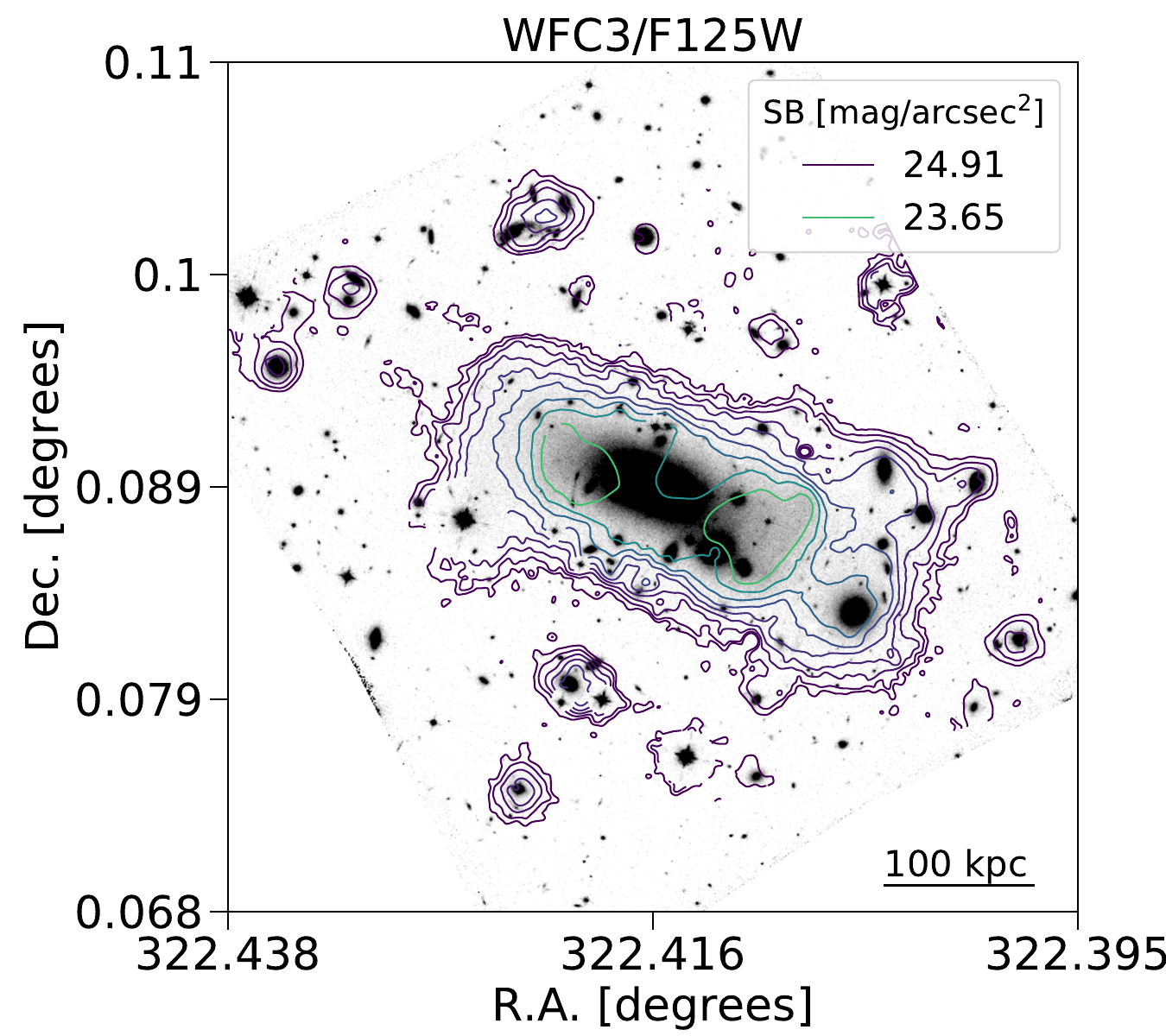}\hfill \\
\includegraphics[width=.48\textwidth]{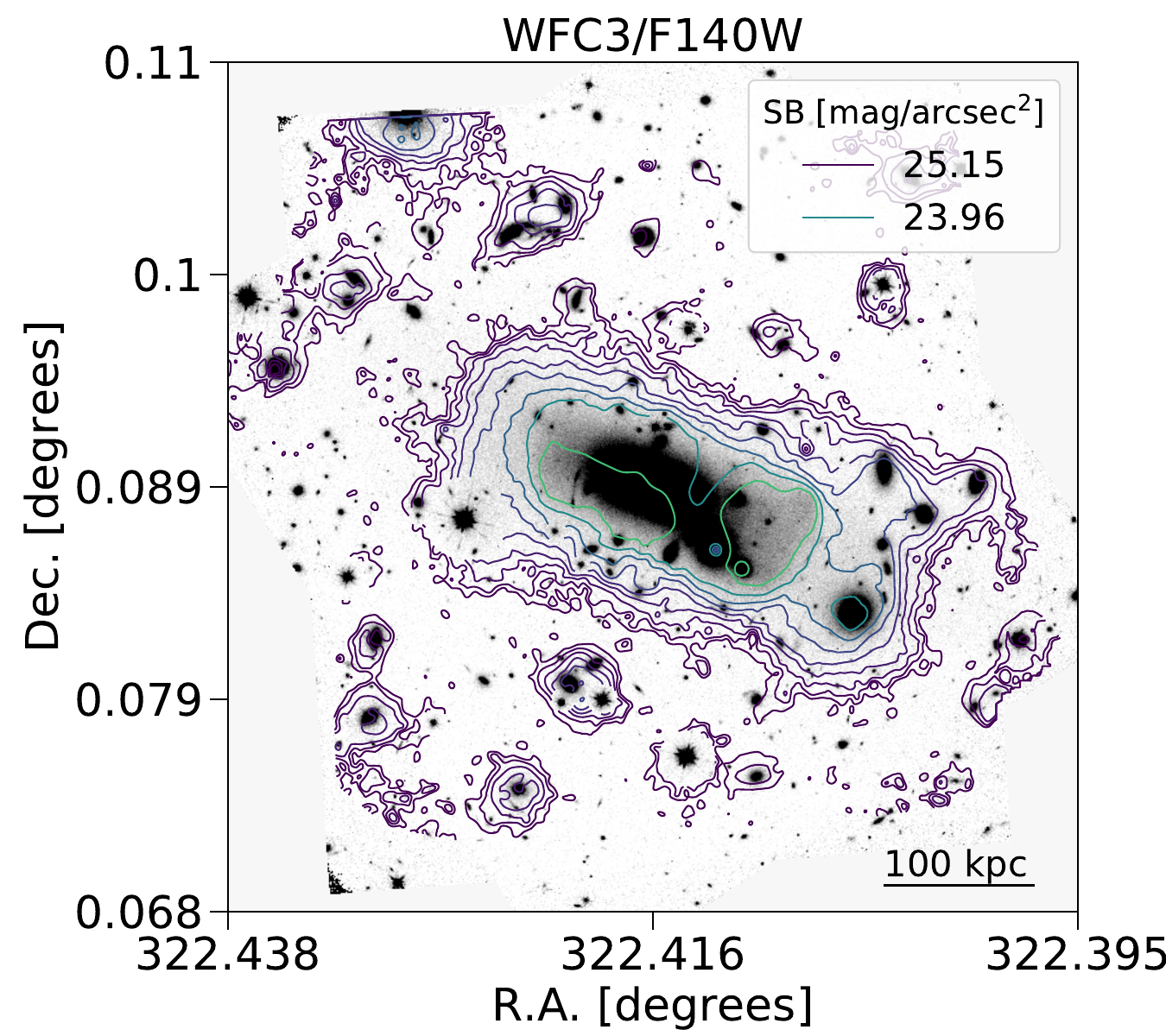}\hfill\includegraphics[width=.48\textwidth]{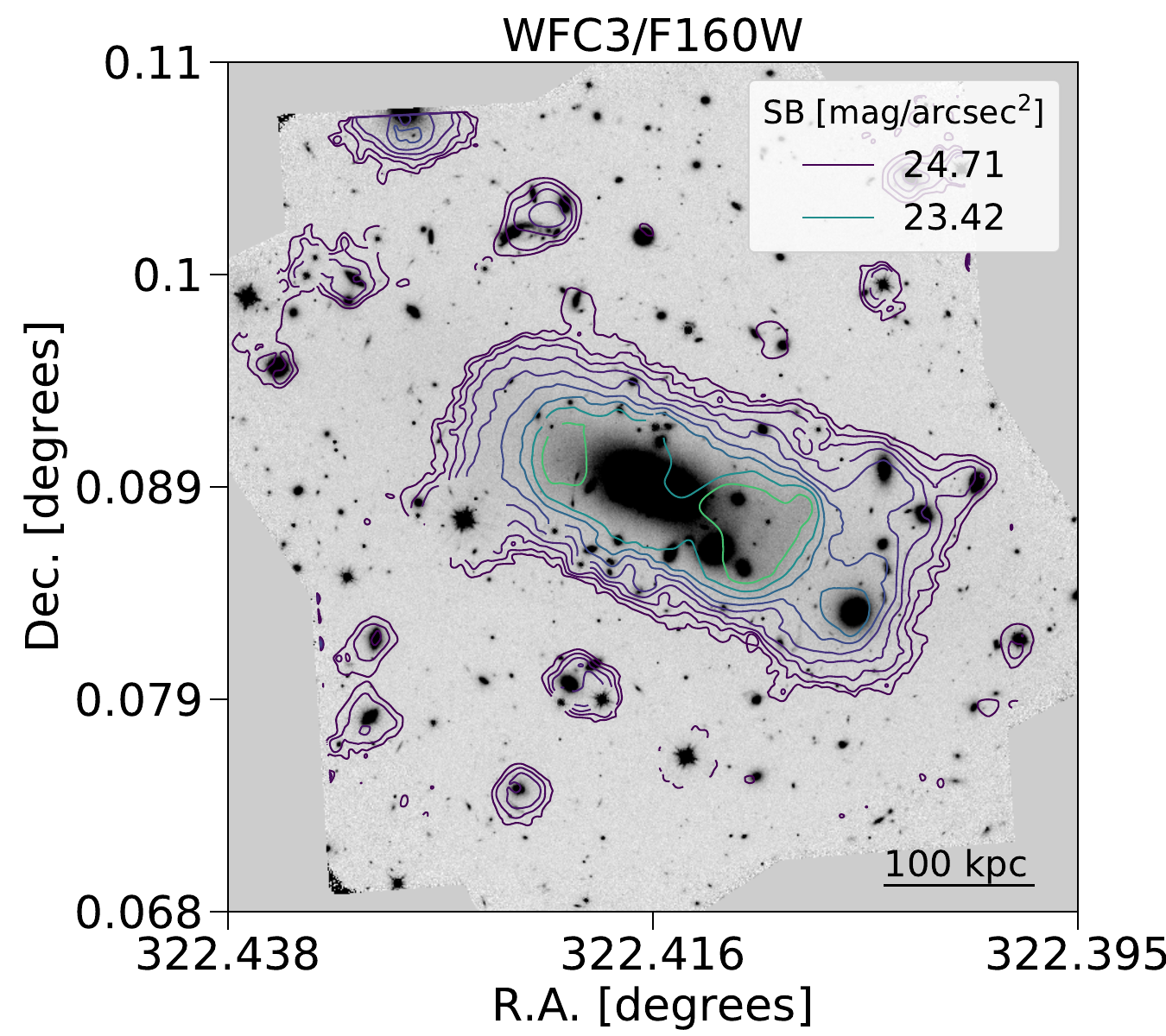}\hfill \\
\caption{ICL isocontours superimposed over the original images in the twelve HST filters considered for this work (continuation of Fig. \ref{fig:ICL_contours_1}).} \label{fig:ICL_contours_2}
\end{figure*}

\begin{figure*}
\centering
\includegraphics[width=.48\textwidth]{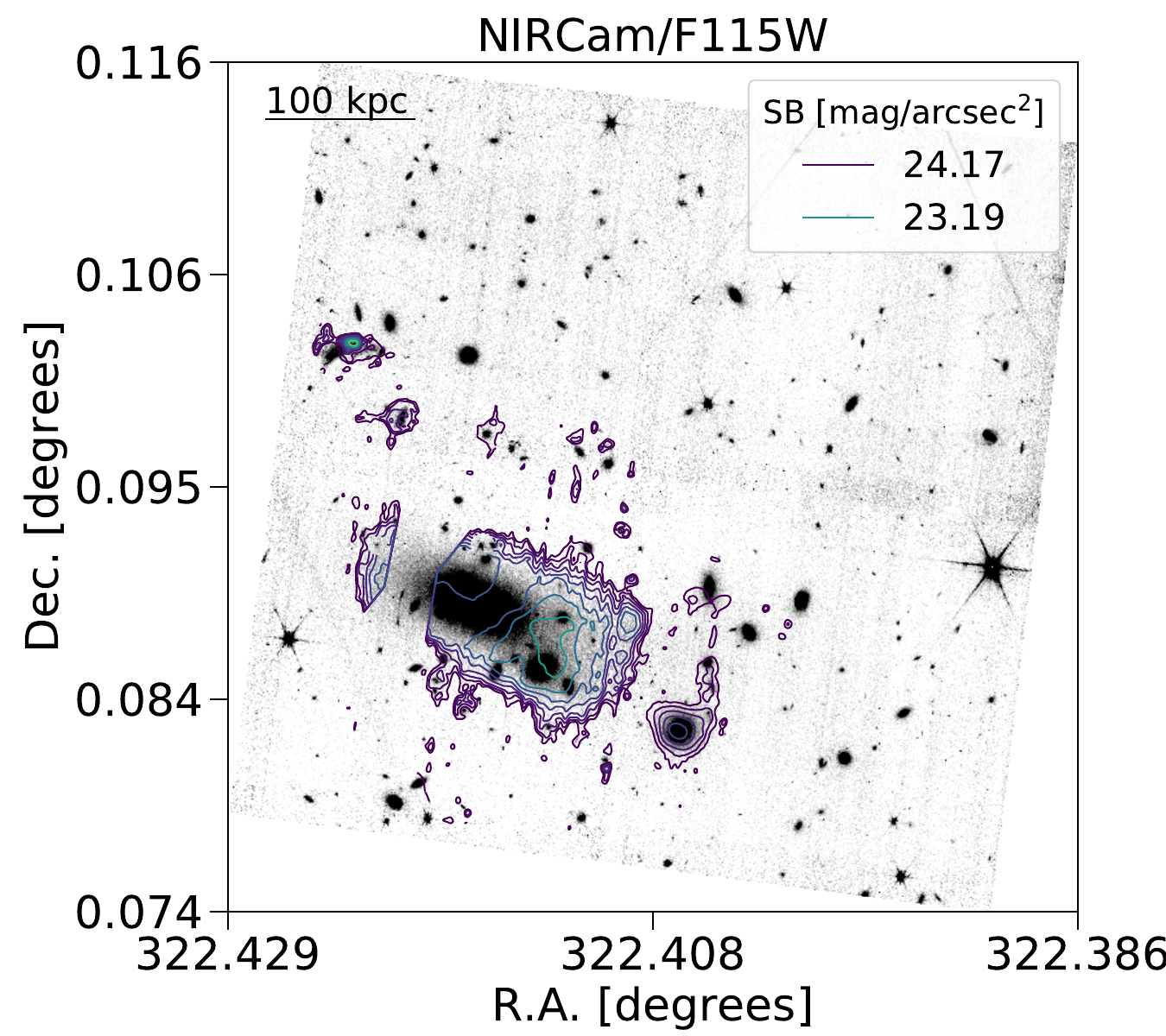}\hfill\includegraphics[width=.48\textwidth]{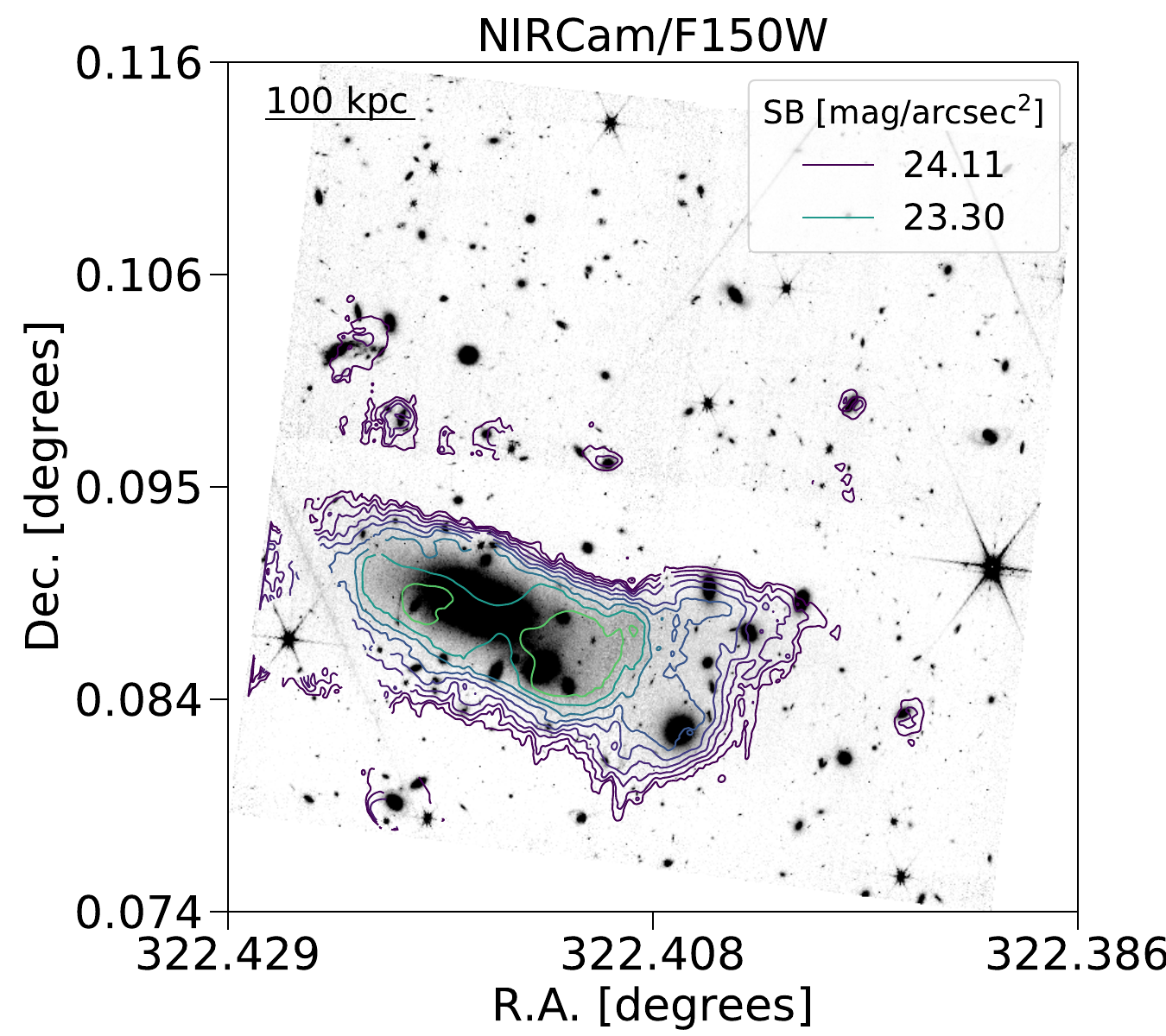}\hfill \\
\includegraphics[width=.48\textwidth]{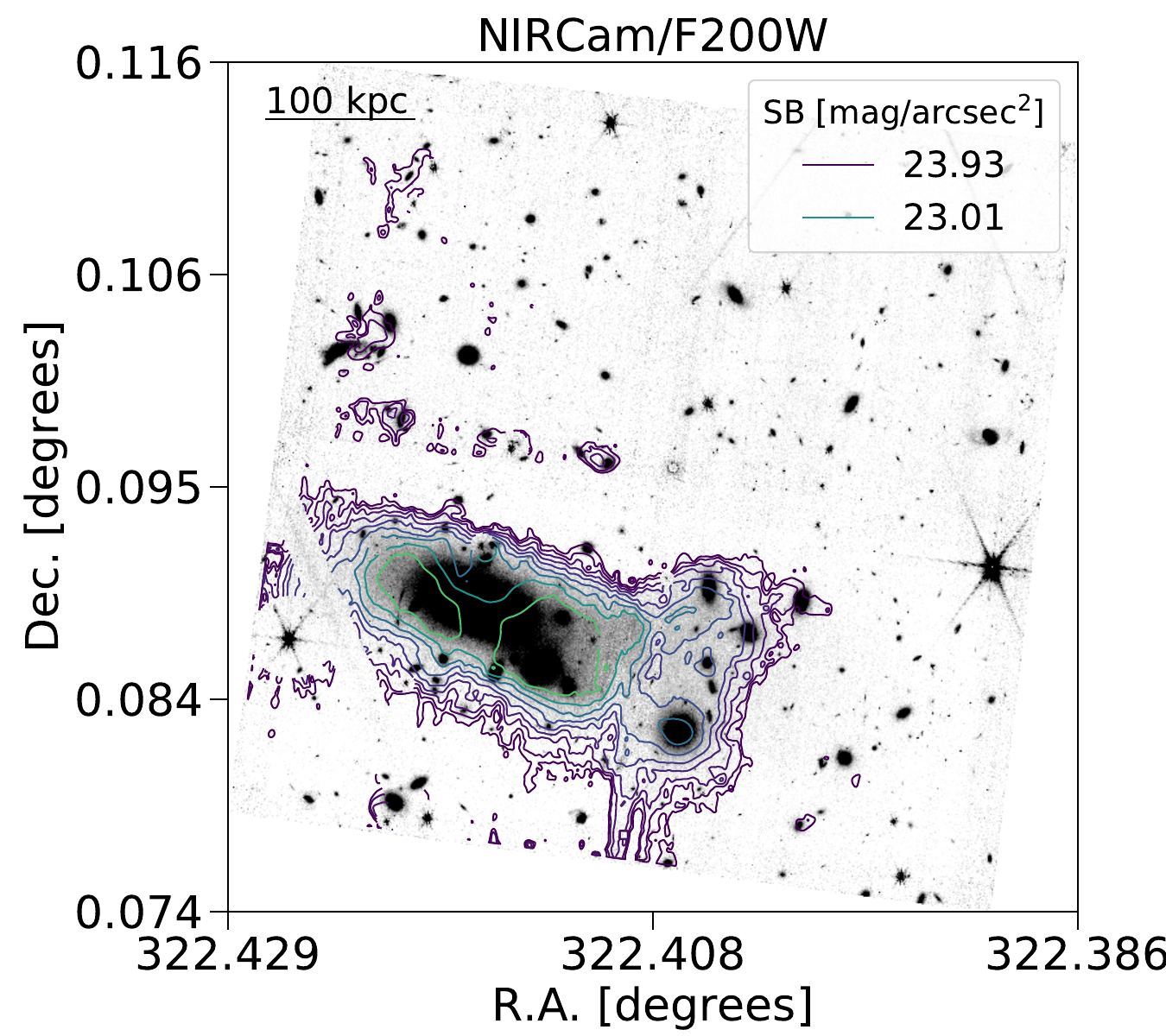}\hfill\includegraphics[width=.48\textwidth]{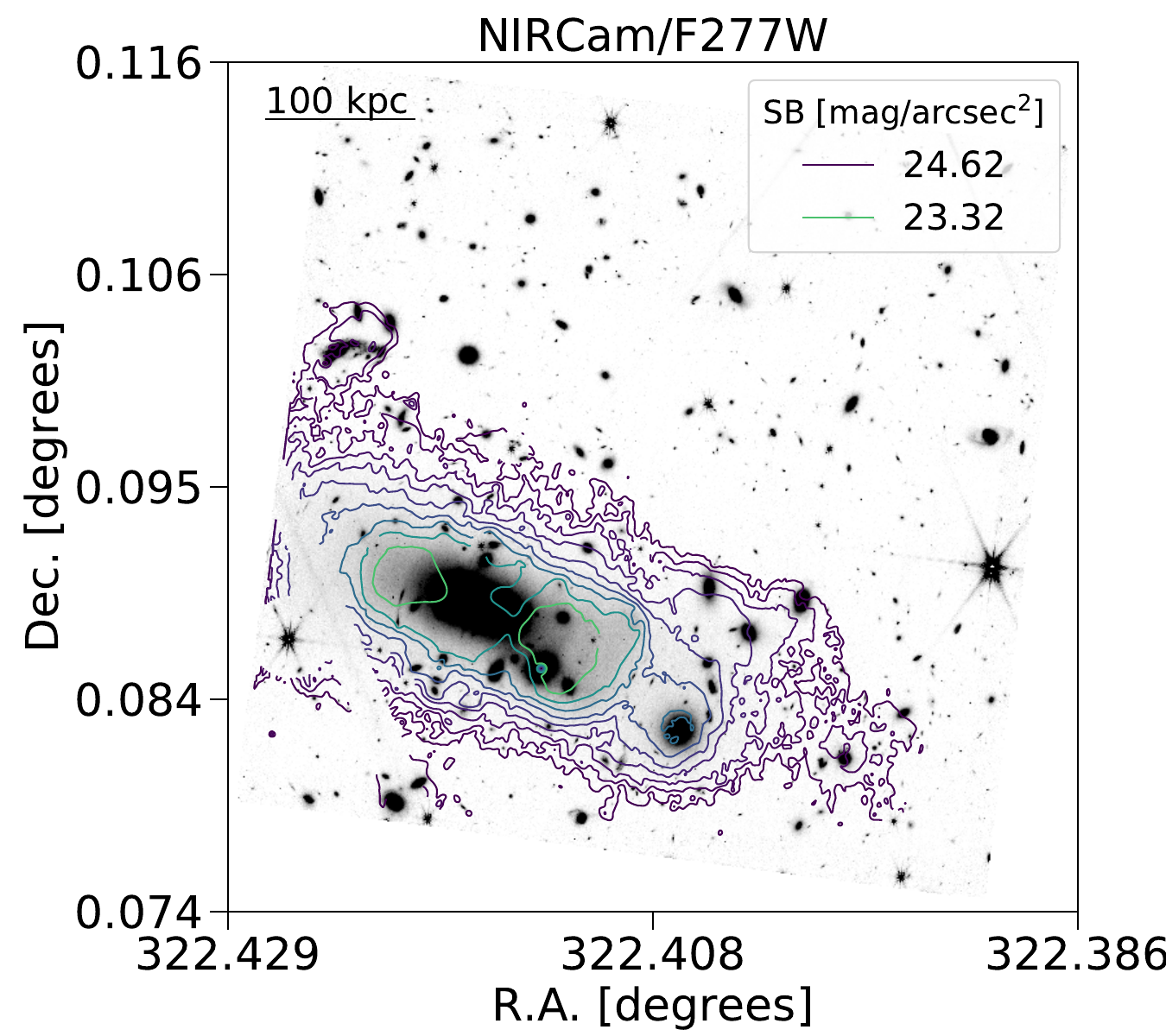}\hfill \\
\includegraphics[width=.48\textwidth]{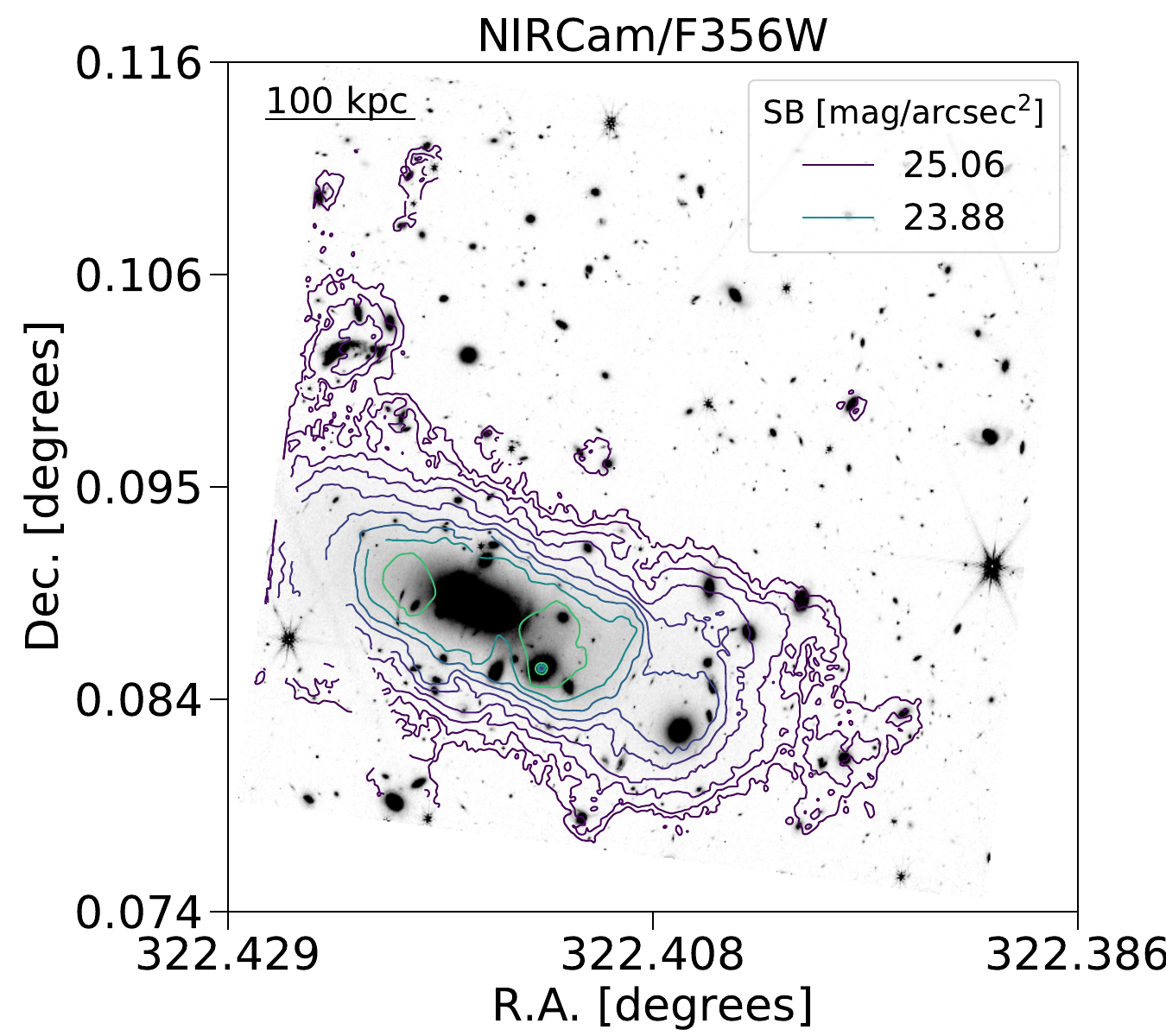}\hfill\includegraphics[width=.48\textwidth]{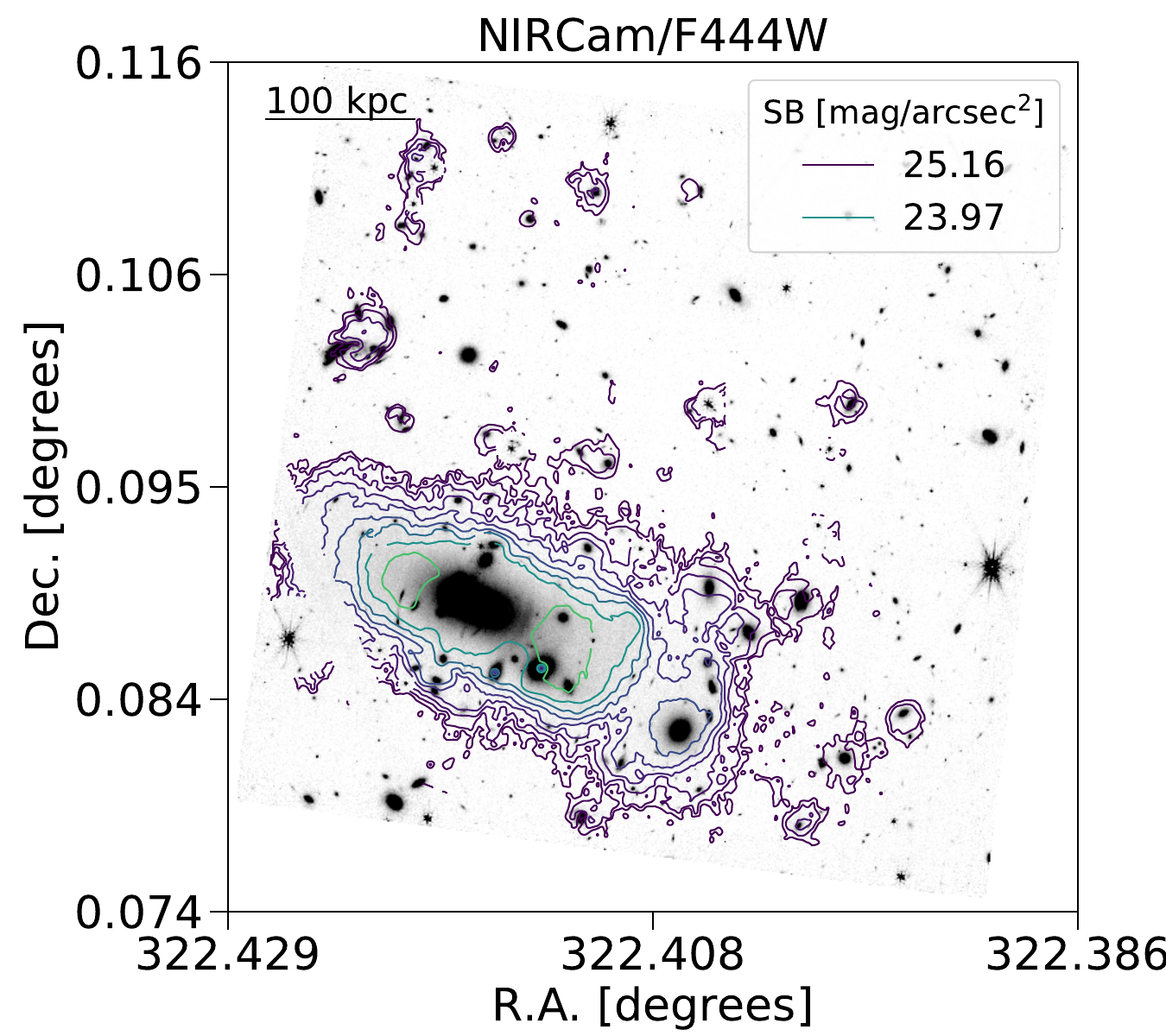}\hfill \\
\caption{ICL isocontours superimposed over the original images in the six JWST filters considered for this work. } \label{fig:ICL_contours_3}
\end{figure*}

We thus ran CICLE independently on all different HST and JWST bands to remove all the galactic light and ended up with 18 images that just contain ICL and sky background. An exquisite estimation of this sky level and its variability across the image are crucial, given the faintness of the ICL. Several recent works have compared different techniques of sky subtraction in the context of low-surface-brightness studies \citep{Borlaff2019,Haigh2021,Kelvin2023}. These include Sextractor \citep{Bertin2002,Bertin2012}, NoiseChisel \citep{Akhlaghi2015,Akhlaghi2019}, and the Vera C. Rubin Observatory \citep{Ivezic2008} and Hyper Suprime Cam Subaru Strategic Program \citep[HSC-SSP,][]{Aihara2018a,Aihara2018b} pipelines. As NoiseChisel is confirmed to be one of the best softwares in this matter, we ran it over the original images to estimate the background maps and remove them from the ICL maps obtained with CICLE. Results can be seen in Figs. \ref{fig:ICL_contours_1}, \ref{fig:ICL_contours_2}, and \ref{fig:ICL_contours_3}, where the ICL isocontours are plotted over the original images, for the twelve HST and six JWST filters considered in this work. We stress here that the ICL maps from the F115W, F150W, and F200W bands will not be used in the subsequent analysis, as we identified some wisps west to the BCG. However, the eastern side of the ICL can be considered clean.\\

We observe a clumpy distribution of the ICL, with a main core surrounding the BCG and other bright companion galaxies, plus several minor concentrations around identified galaxy members. These patches of ICL have a considerably lower level of significance compared with the main core, as it can be inferred from the surface brightness levels of the isocontours. We do not observe a significant difference in the ICL morphology from the bluest to the reddest filters: the main core of the ICL is well aligned with the BCG (main axis with a position angle of $\sim 70^{\circ}$) and other bright cluster members and it is preferentially concentrated around two peaks, east and west of the BCG respectively. This main cloud is fairly symmetric, although this symmetry is somehow broken by the larger extension of the secondary cloud at its western side. It is important to remark that this secondary cloud appears at all wavelengths, and it can appear detached from the main cloud in the intermediate-wavelength images (from F606W to F850LP). The western side of the secondary cloud is primarily concentrated around a bright cluster member located on its southern side, but it spreads out to the northwest as we go towards redder wavelengths.  \\

Interestingly, a lower-significance extension of the ICL is also visible in the F277W and F356W filters, spreading outwards to the southwest following the same position angle of $\sim 70^{\circ}$ mentioned before. This may be due to the higher signal-to-noise of the JWST LW channels, given their coarser native pixel scale with respect the SW filters, or due to the red color of the ICL. The finer substructure identified by eye in the false color image of RXJ2129 (Fig. \ref{fig:RGB_original} right), is now evidenced by the isocontours in the different filters. The main cloud isocontours are more boxy west to the BCG and have a clear protuberance in the region of the clump. In Fig. \ref{fig:regions}, we show zoomed-in false color images of both the cluster {(top)} and the ICL {(bottom)}, generated from the IR HST/WFC2 and the JWST/NIRCam/LW images. We have explicitly decided to keep the compact sources (possibly globular clusters) in the output ICL maps, as they are likely part of the intracluster material, although CICLE is able to remove them too. In the following sections, we will study each one of the features and substructures found in the ICL with more detail.\\

\section{Generation of the ICL maps: MUSE} \label{sect:ICLmaps_MUSE}

Although the MUSE-DEEP science products provided by ESO already have the sky subtracted, in the case of RXJ2129 we observed that some sky lines were still present. We used the Zurich Atmosphere Purge \citep[ZAP, ][]{soto2016}, a Python module specially designed to remove the background in MUSE datacubes. ZAP can be applied to previously sky-subtracted cubes, as it was our case, enhancing the existing subtraction by isolating the residual sky features with a principal component analysis. We then ran CICLE iteratively over the clean datacube with the same configuration as that of the HST and JWST images, allowing a maximum of 15 components for each polar coordinates. We finally obtained 3681 ICL maps spanning the wavelength range $\sim 475-935$ nm. \\

\section{SED fitting} \label{sect:SEDfitting}

\begin{figure*}
\centering
\includegraphics[width=0.6\textwidth]{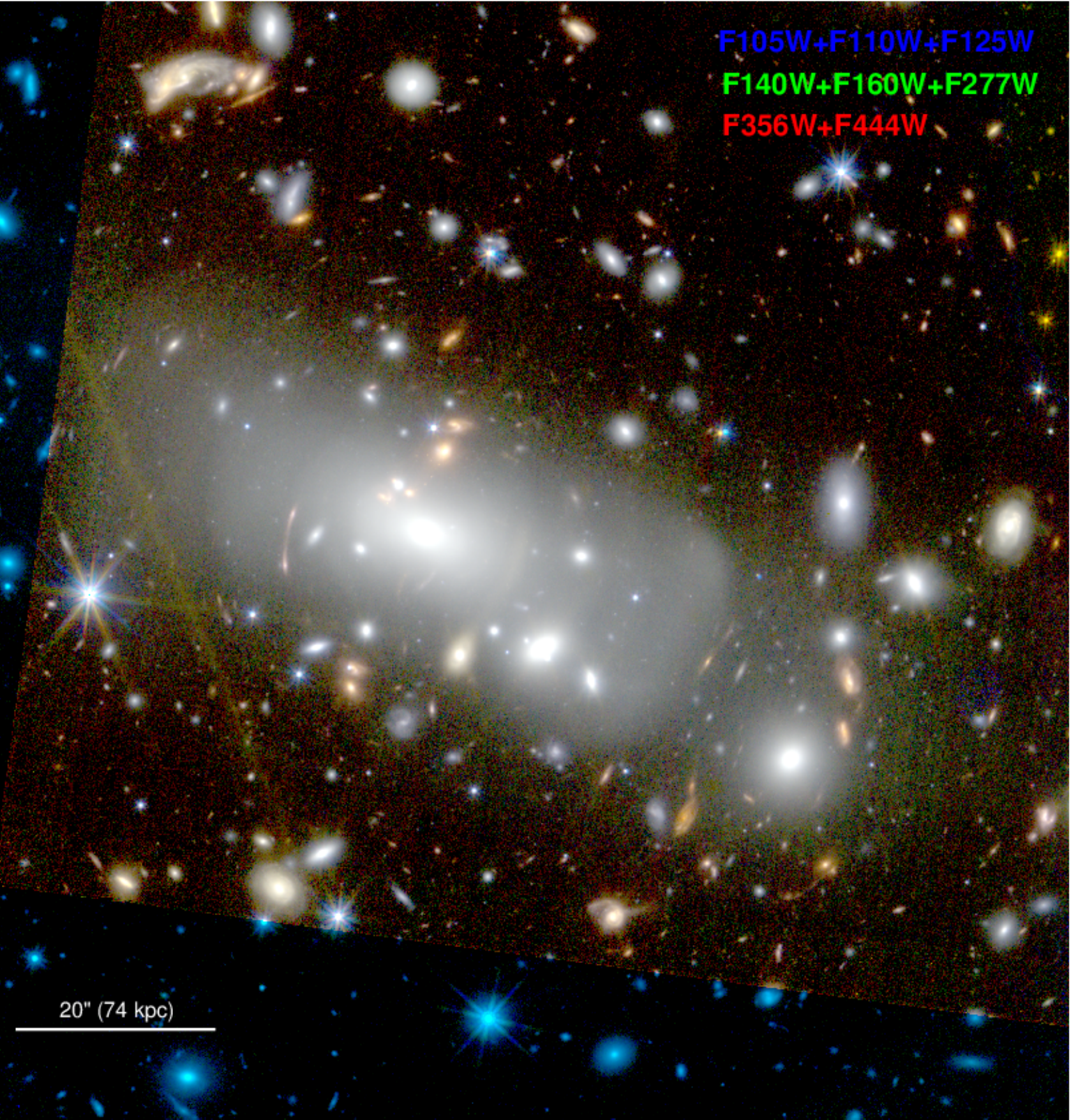}\\ 
\includegraphics[width=0.6\textwidth]{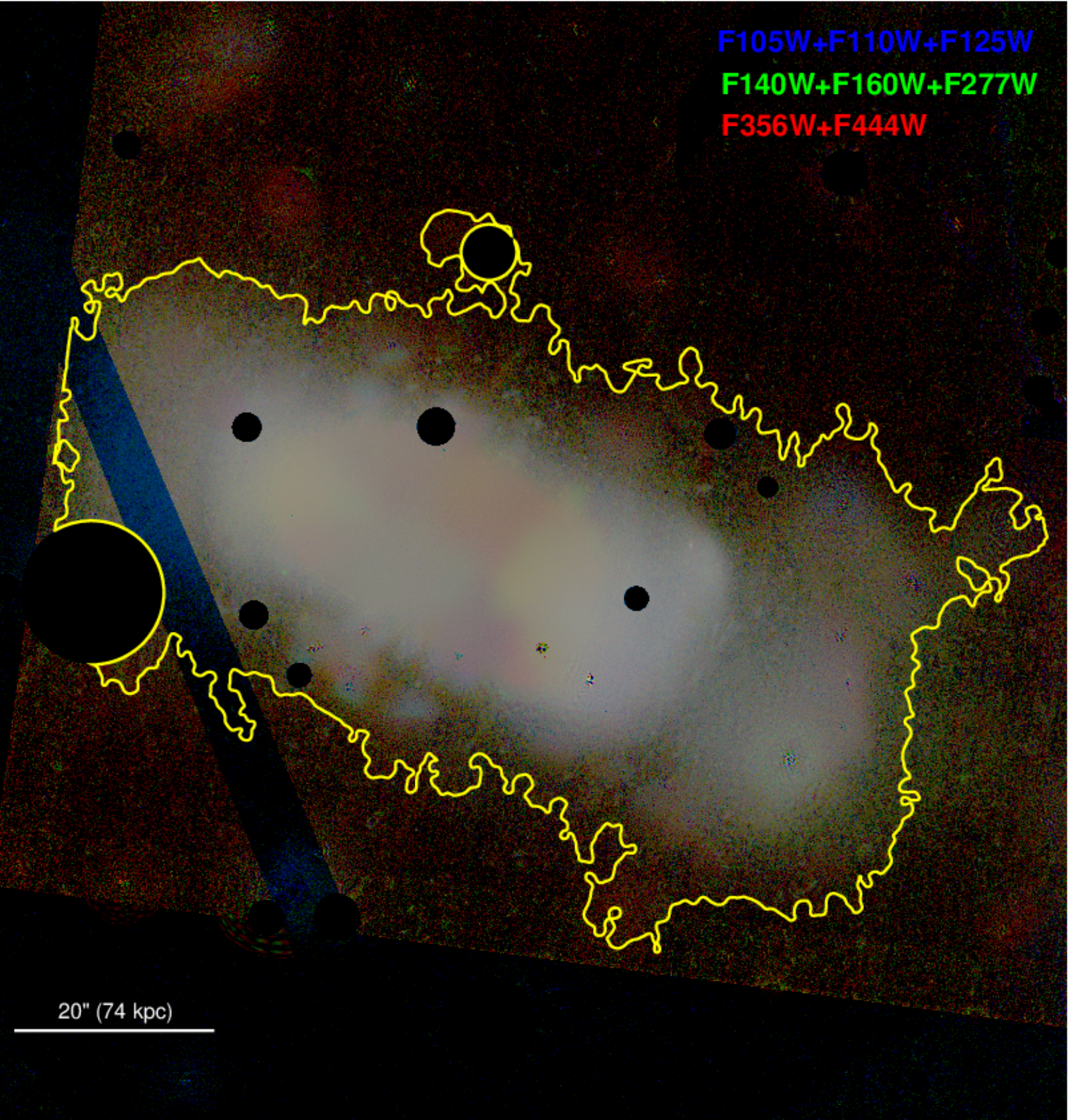}
\caption{Infrared ICL emission in RXJ2129. Zoomed-in false color image of RXJ2129 built from the IR HST and JWST bands (top) and the corresponding ICL calculated with CICLE (bottom). Both color images were made with Trilogy and are at the same scale. Dark circles and blue diagonal band in the ICL map are masked pixels, corresponding to stars and diffraction spikes. The yellow line outlines the total region where we measure the photometry.} 
\label{fig:regions}
\end{figure*}

\begin{figure*}
\centering
\includegraphics[width=\textwidth]{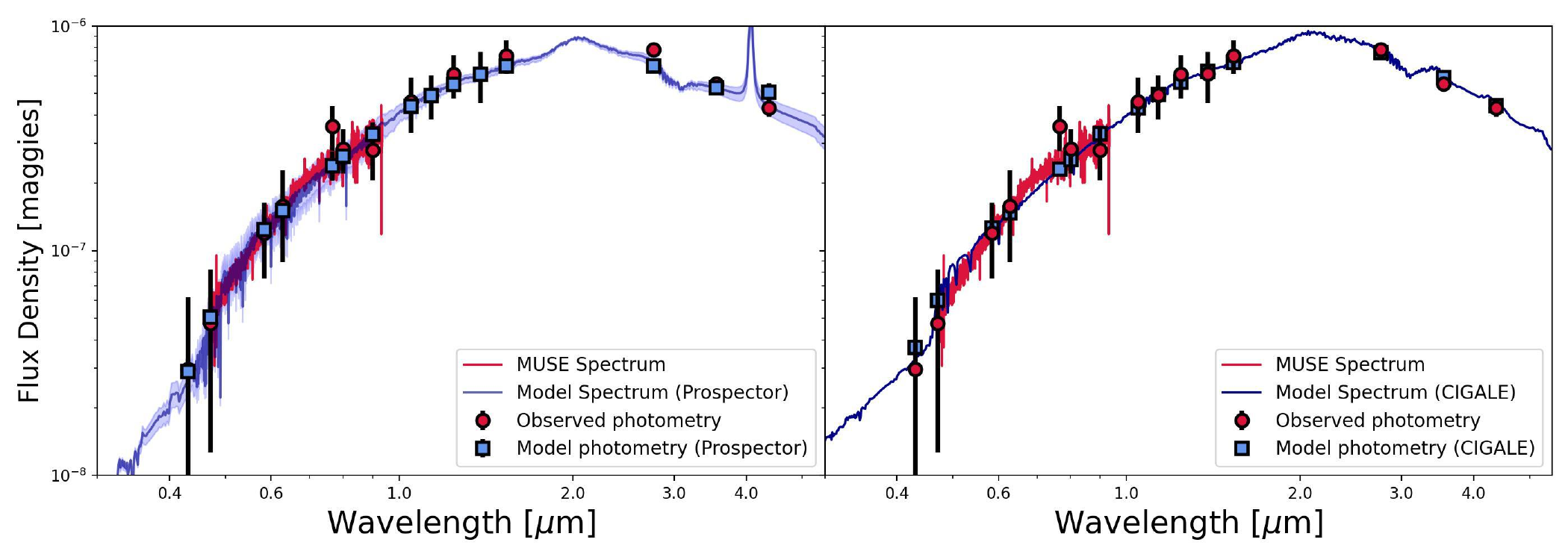}
\caption{ICL SED fits obtained with Prospector (left) and CIGALE (right). The data are plotted in red: a line for the MUSE spectrum an filled circles for the HST+JWST photometry. The result of the fit is represented in blue: a line for the best fit model and filled squares for the expected photometry based on the model. } \label{fig:SEDfitting}
\end{figure*}

We analyze the properties of the ICL stellar populations by fitting its 15-band spectral energy distribution (SED) simultaneously with the optical spectra in different regions. We used the JWST F444W image to outline the whole extension of the ICL, using the isocontour with the lowest surface brightness (see Fig. \ref{fig:ICL_contours_3} bottom right). The total region comprises the main and secondary clouds. Due to its lower significance, we leave the southwest extension out of this region (see Fig. \ref{fig:regions}). We measured the flux within this region in the HST and JWST imaging and the MUSE datacube. The HST and JWST photometry measured is listed in Table \ref{table:observations}, along with the corresponding errors. These errors are the combination of three different sources of uncertainty: the background noise, the error in the aperture sum, and the Poisson noise. The MUSE spectrum and the broad-band SED are shown in Fig. \ref{fig:SEDfitting}. Further analyses to fit the SEDs of the clump and the shells, along with other local regions of interest, will be the scope of a forthcoming paper. \\ 

The optical to near IR spectral coverage, given by the combination of HST and JWST data, allows an accurate modeling of the SED. Additionally, the MUSE spectra {provide} an exceptional spectral resolution to study the ICL stellar populations. In order to compute age and mass of the stars that form the ICL we employed two different tools of SED fitting: CIGALE \citep[Code Investigating GALaxy Emission; ][]{boquien2019} and Prospector \citep{johnson2021}, two state-of-the-art Bayesian codes.  CIGALE uses parametric star formation histories (SFHs) and determines the output physical parameters and their uncertainties by using a Bayesian approach, i.e., it creates a probability distribution function for each parameter by evaluating the $\chi^{2}$ over the full set of models used for the fit. The final values of the output parameters are derived as the mean of the probability distribution function and the corresponding uncertainty is the standard deviation. The modules used are a delayed star formation history with optional exponential burst, stellar population models by \cite{bruzual2003}, and a \cite{Chabrier2003} initial mass function. The attenuation in the models was taken into account by using the modified \cite{Calzetti2000} law. Differently, Prospector incorporates non-parametric SFHs \citep[Continuity Flex SFH; ][]{leja2019} and a sophisticated dust model \citep[e.g., ][]{lower2022}. Remarkably, Prospector fits all parameters at once with Markov Chain Monte Carlo or nested sampling to produce joint high-dimensional constraints. The final values are derived as the mean of the posterior probability distribution, with uncertainties determined by the 16th and 84th percentiles. It is important to remark that Prospector allows to combine both spectroscopic and photometric data to better constrain the parameters space, while CIGALE uses photometric information solely.\\

Although this approach is more common to fit the SED of galaxies, for which the SFH is well defined, it is still valid for the ICL. The ICL formation process (stars that have been gravitationally stripped from their parent galaxies through cluster galaxies interacting with other galaxies or with the cluster potential) is  completely unbiased. Older galaxies predominantly contribute with older stars while young stars come from in situ star formation or younger galaxies and the outskirts of massive galaxies. Consequently, the resulting ICL stellar population statistically represents a SFH of the entire system.\\

As shown in Fig. \ref{fig:SEDfitting}, the resulting models nicely fit the observed data, both the photometric measurements and the MUSE spectrum. Moreover, the models and physical parameters estimated are consistent between both SED fitting codes. Indeed, both Prospector and CIGALE models exhibit a similar trend, rising and peaking at approximately 1.6 $\mu$m, followed by a gradual decline. The Prospector SED reproduces the Balmer Break at 0.4 $\mu$m, a feature not observed in the CIGALE fit. Additionally, the Prospector SED shows a peak near 4.4 $\mu$m, likely attributable to Polycyclic Aromatic Hydrocarbon (PAH) emission associated with CN, which may influence the photometric measurements at this wavelength. It is important to notice that the bluest part of the SED is not well constrained by our data points. Prospector presents a more attenuated spectrum in this region compared to CIGALE. However, this difference does not affect the optical and near infrared regions, where both fits are similar. Both SED models were constructed without including nebular emission modules.\\

CIGALE estimates an ICL stellar mass of  $log_{10}(M^*M_{\odot}) = 11.81 \pm 0.34$ and a main age of the older population of $10.49 \pm 0.41$ Myr. Prospector, accordingly, yields a total mass of $log_{10}(M^*/M_{\odot}) = 12.33\pm 0.34$ with the majority of stars formed in an age bin of 9.7 to 10.35 Myr (see Table \ref{table:SEDfitting_physicalproperties}). For both codes, the age reported refers to the older population. Although metallicity is less well constrained, the values obtained by both codes are consistent with being solar.\\ 

\begin{table}
\centering
\begin{tabular}{l|ccc}
 Method & $log_{10}(M^*/M_{\odot})$ & Age & $log_{10}(Z/Z_{\odot})$ \\
 & & [Gyr] \\
 \hline
 CIGALE & $11.83 \pm 0.14$ & $10.49 \pm 0.41$ & $-0.77 \pm 0.31$\\
 Prospector & $12.33^{+0.04}_{-0.03}$ & [9.7,10.35] & $-0.51 ^{+0.22}_{-0.20}$\\
\end{tabular}
\caption{Physical properties of the ICL stellar populations, as derived by CIGALE and Prospector algorithms: stellar mass, main age of the older population, and metallicity.} \label{table:SEDfitting_physicalproperties}
\end{table}

\section{X-ray analysis} \label{sect:x-rays}

In order to produce an exposure corrected image  for the analysis of the region of interest only the energy interval from 0.5 keV to 7 kev was used. Aspect histograms were created with the task \textit{asphist} and instrument maps were generated with the task \textit{mkinstmap}. Exposure maps were created with the tool \textit{mkexpmap}, where the dimensions were determine with the task \textit{get\_sky\_limits}. The image was normalized by the exposure maps using the standard \textit{dmimgcalc}. \\ 

The tool \textit{specextract} was used to produce spectral files and responses for the regions analyzed. Spectra was grouped to have 15 cnt/channel. Given the small region of interest near the cluster center, we used a local background region in the same CCD, away from the cluster’s center about 1.5 Mpc, which is comparable to the cluster's R$_{200}$ estimated from its X-ray measured characteristic temperature \citep{monique99}. Spectral fittings were carried out  in \textit{XSPEC 12.13.0c},  using an absorbed Collisional Ionization Equilibrium model \textit{phabs*apec}, with the  redshift 0.234 fixed at their nominal value of the BCG and the Hydrogen column density (nH) of 3.4$\times10^{20} cm^{-2}$ chosen from the HI4PI Map \citep{HI4PI} through the HEASARC \textit{nH} tool\footnote{heasarc.gsfc.nasa.gov/cgi-bin/Tools/w3nh/w3nh.pl}.\\

\begin{figure*}
\centering
\includegraphics[width=\hsize]{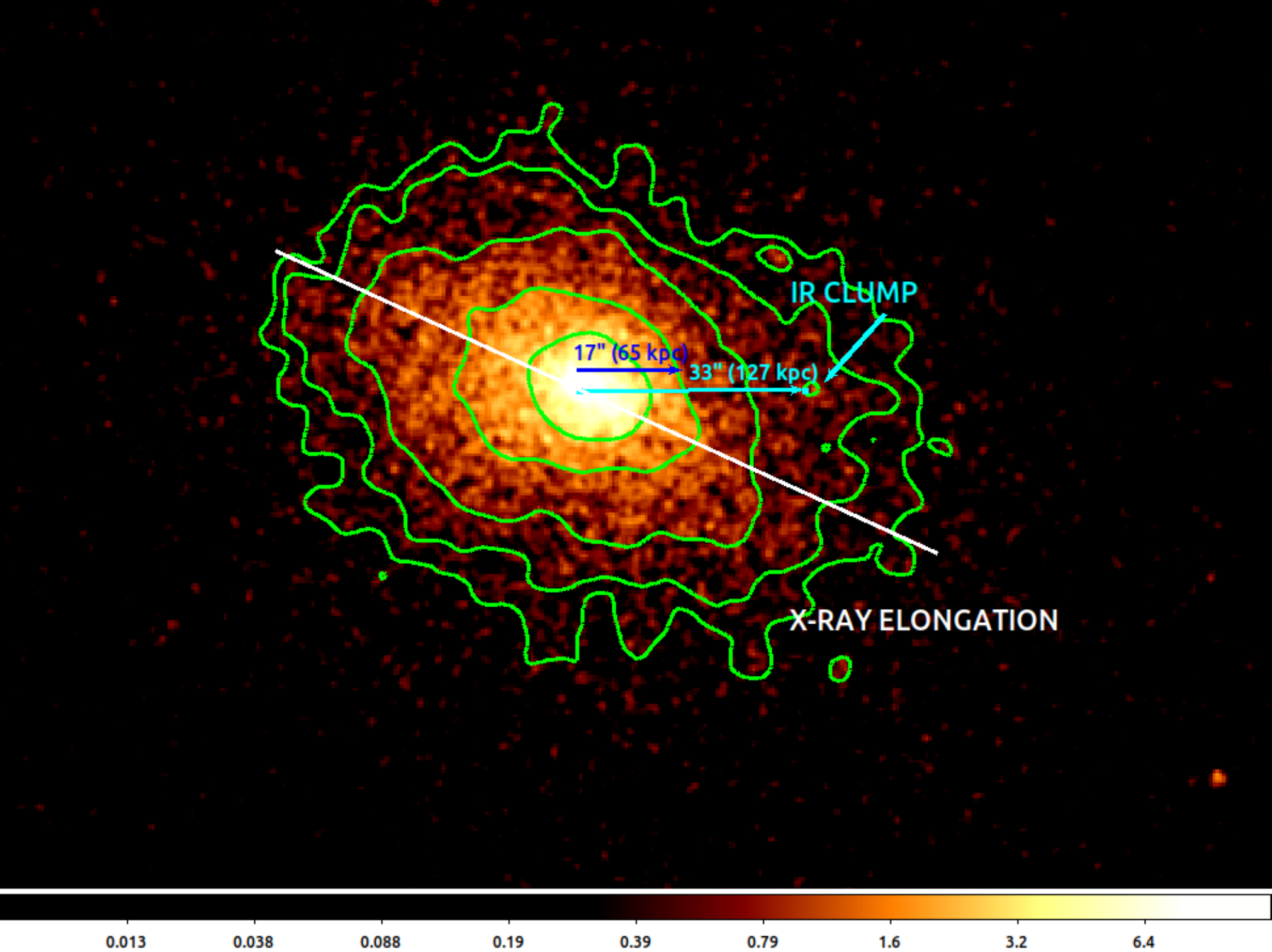}
\caption{{Chanda X-ray image of RXJ2129. Its isocontours show a similar elongation as the BCG and ICL. Excess X-ray emission can be seen clearly in the location of the IR clump to the west. A surface discontinuity is also seen in the contour level indicated by the dark blue arrow. Distances from the BCG center to the surface discontinuity and to the IR clump region are also shown on top of the dark blue and cyan arrows, respectively. $1”$ corresponds to 3.75 kpc.}}\label{x-cont}
\end{figure*}

\begin{figure}
\centering
\includegraphics[width=\hsize]{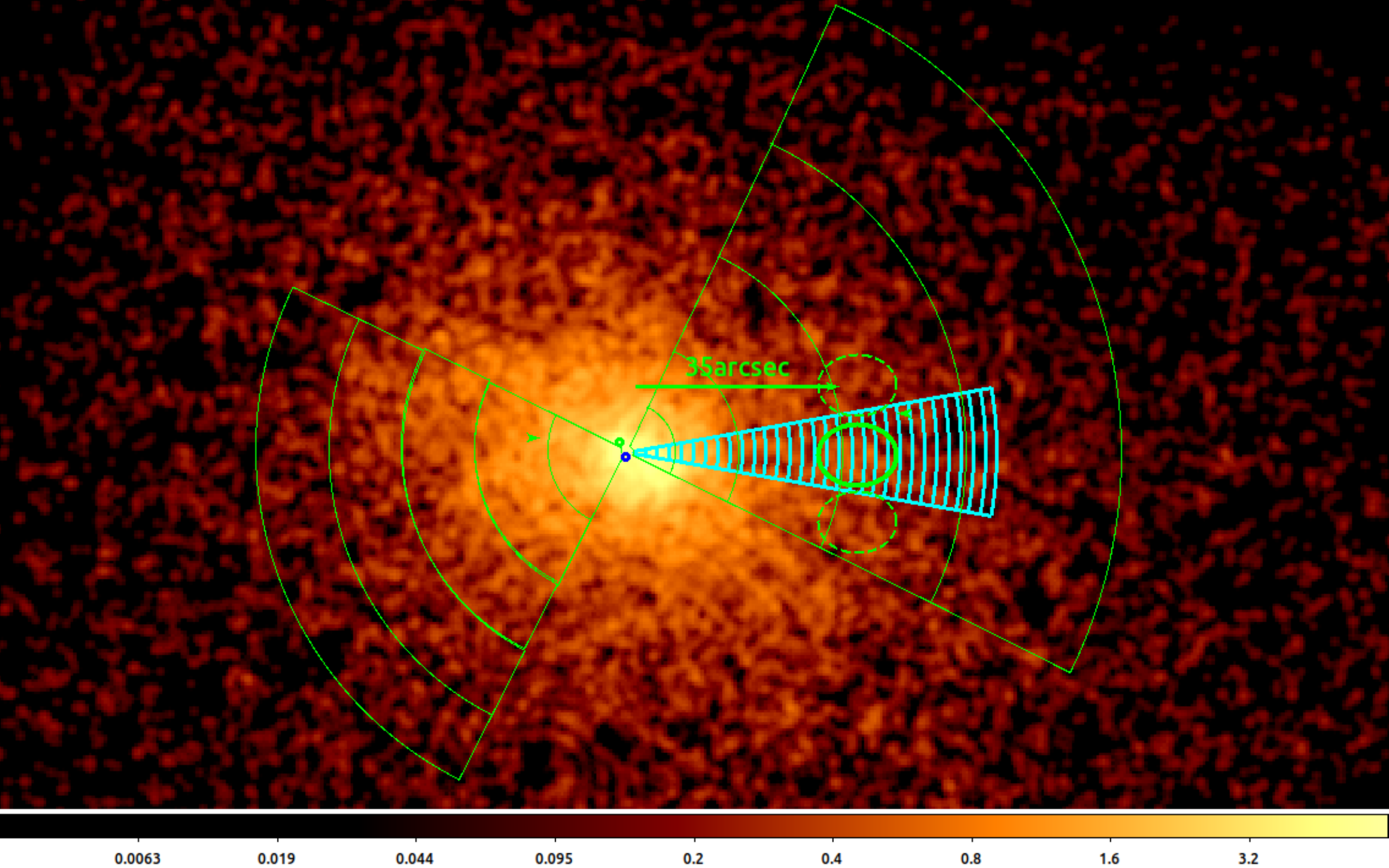}\\
\caption{Extraction regions for imaging (thin) towards the ICL/X-ray clump (green ellipse) and temperature distribution (broad) towards and away from the SB discontinuity. The two nearby regions to measure the azimuthal excess are also shown (dashed)}\label{extraction}
\end{figure}

\begin{figure*}
\includegraphics[width=0.48\textwidth]{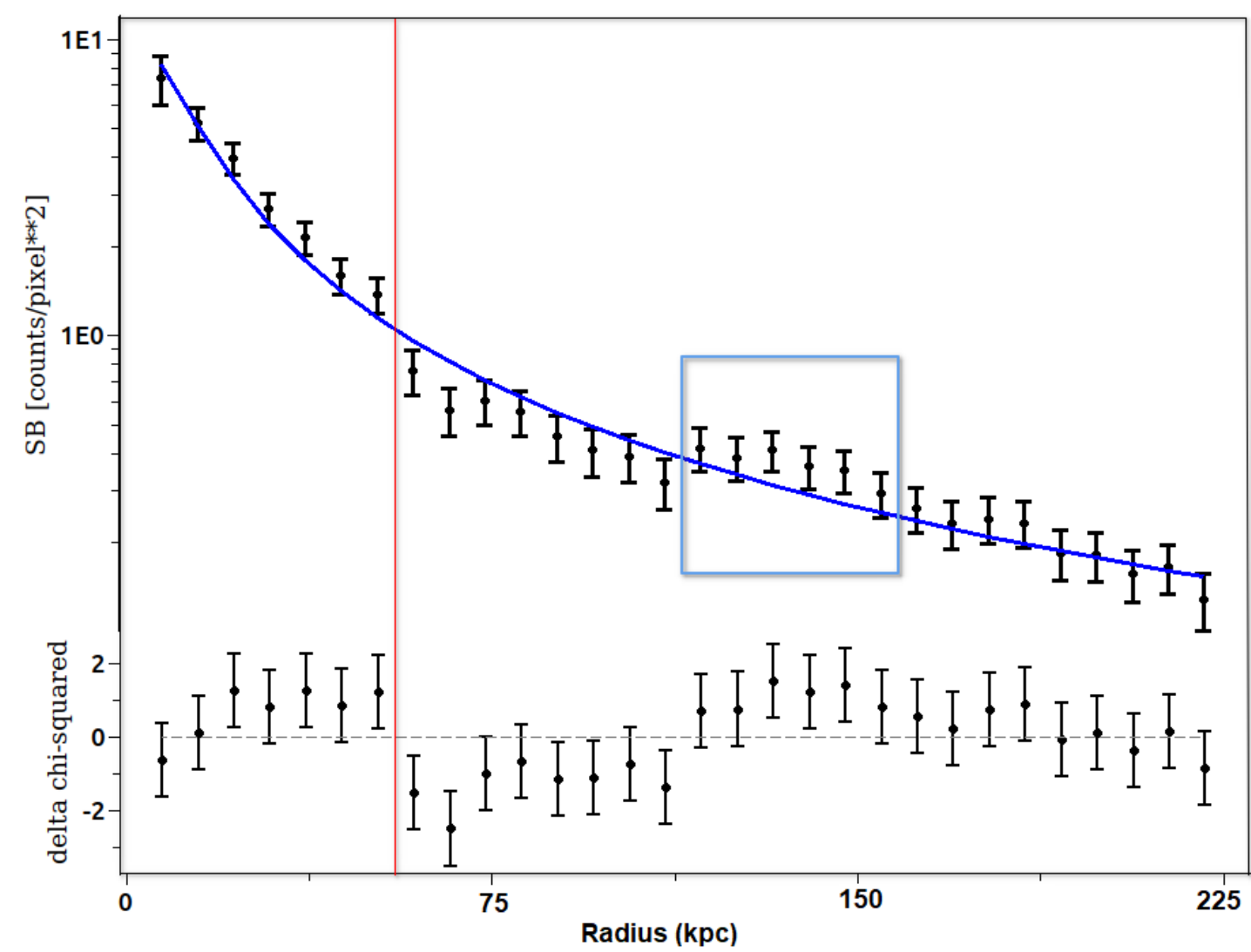}
\includegraphics[width=0.49\textwidth]{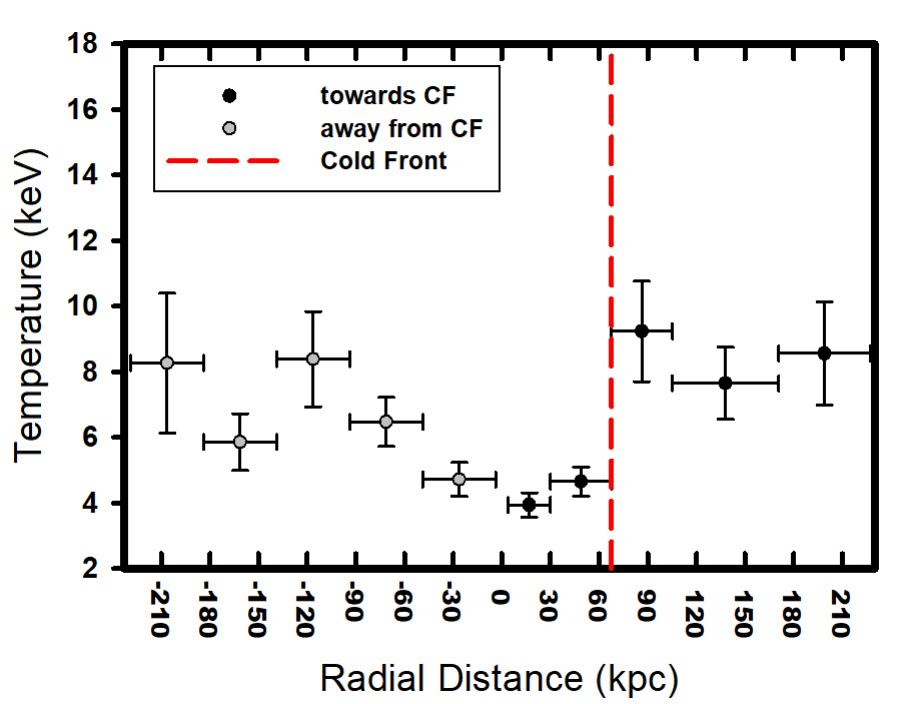}
\caption{{Left: X-ray SB and residuals from a single beta model fit in the direction of the clump. The red vertical line indicates the position of the surface brightness discontinuity. The light blue box shows the region boundaries of X-ray emission excess in the location of the clump. Right: Projected temperature profile along E-W direction near the center of the cluster. The surface brightness discontinuity is shown by the dashed red line.}}
\label{radial5}
\end{figure*}

The Chandra image of RXJ2129 shows a few interesting features. The X-ray emission shows a clear elongation in the same position angle as that of the BCG+ICL system along an NE-SW direction, namely $\sim 65^{\circ}$ (Fig. \ref{x-cont}). A few {inhomogeneities} are seen in the X-ray image and, among them, a sharp surface brightness discontinuity at about 65 kpc from the center towards the W-NW. To explore the nature of this discontinuity, we extracted the surface brightness profile using the configuration shown in Fig. \ref{extraction} in cyan. We plot the results in Fig.\ref{radial5} left, along with a 1D-$\beta$-model fit and its residuals as a reference. The excess emission at about 33\arcsec~ ($\sim$ 127 kpc) is interestingly close to the clump region. We can also see the sharp decline in surface brightness of the region where the discontinuity is found in a larger scale in {Fig. \ref{x-cont}}. This type of surface brightness discontinuity is not unusual near the center of clusters and can be due to cold fronts originated from the residual movement of the core (sloshing) after a previous off-center merger \citep[e.g.,][]{ascasibar,dupkea496}, or due to a current merger of two clusters, typically accompanied by a shock front upstream \citep[e.g.,][]{a3667}. Given that the two mechanisms for creating sharp discontinuities correspond to very different dynamical states, it is worth checking what kind of discontinuity this is.\\

We extracted the intracluster gas temperature profile, obtained through spectral analysis. Given the short observation available, we used larger radial bins to include $\gtrsim$ 700 counts, delineating the areas of interest, in average with 800 counts per region (Fig. \ref{extraction}). The results are listed in Table \ref{bestfit_proj} and plotted in Fig. \ref{radial5} right, which shows the temperature profile towards (in black) and away from the discontinuity (in grey). A clear jump in temperature is seen at the surface brightness discontinuity. The temperature goes from $\sim 4.7\pm0.5$keV to $9.3\pm 1.5$ keV. The pressure ratio of the regions inside and outside of the discontinuity is found to be $0.83\pm0.17$. The continuity of the pressure profile with the abrupt change of density downwards and temperature upwards strongly suggests that this is indeed a cold front. \\

\begin{figure*}
 \resizebox{6cm}{!}   {\includegraphics[width=\linewidth]{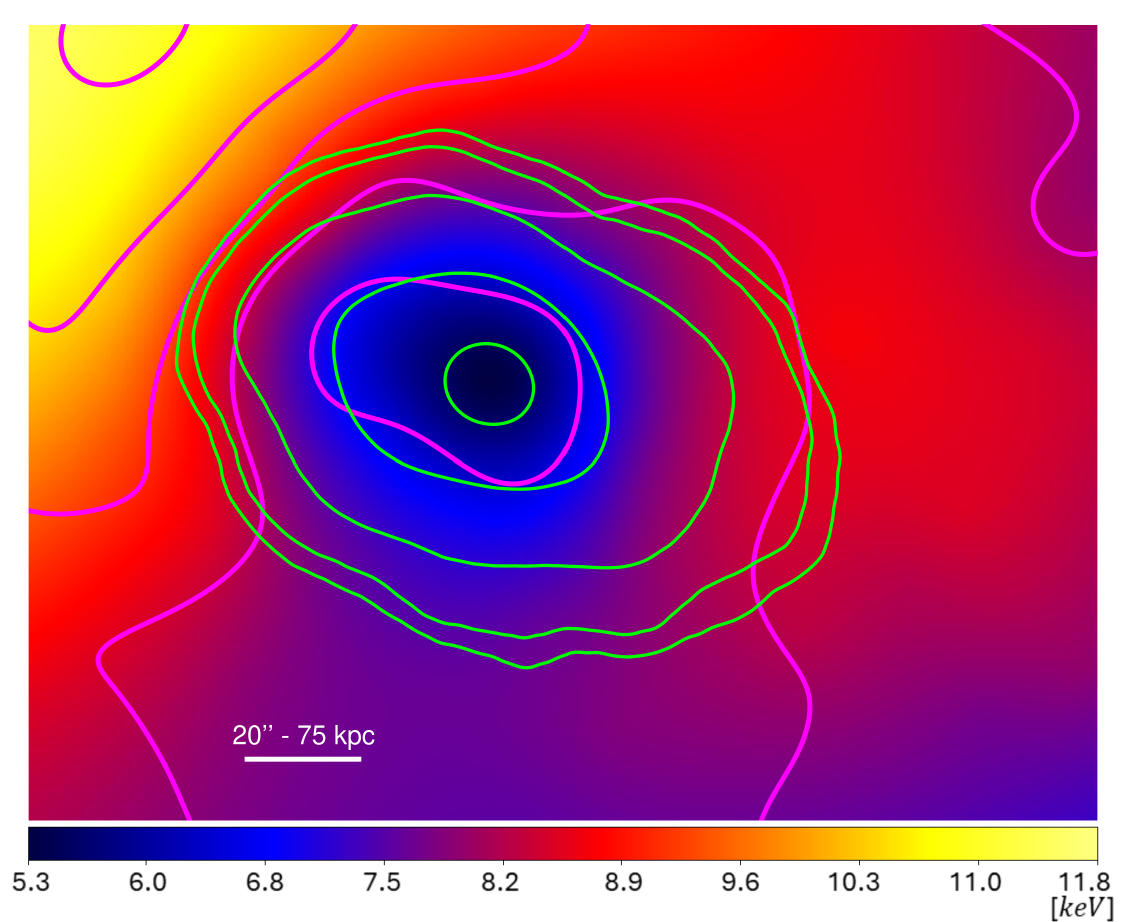}}
\resizebox{6cm}{!}     {\includegraphics[width=\linewidth]{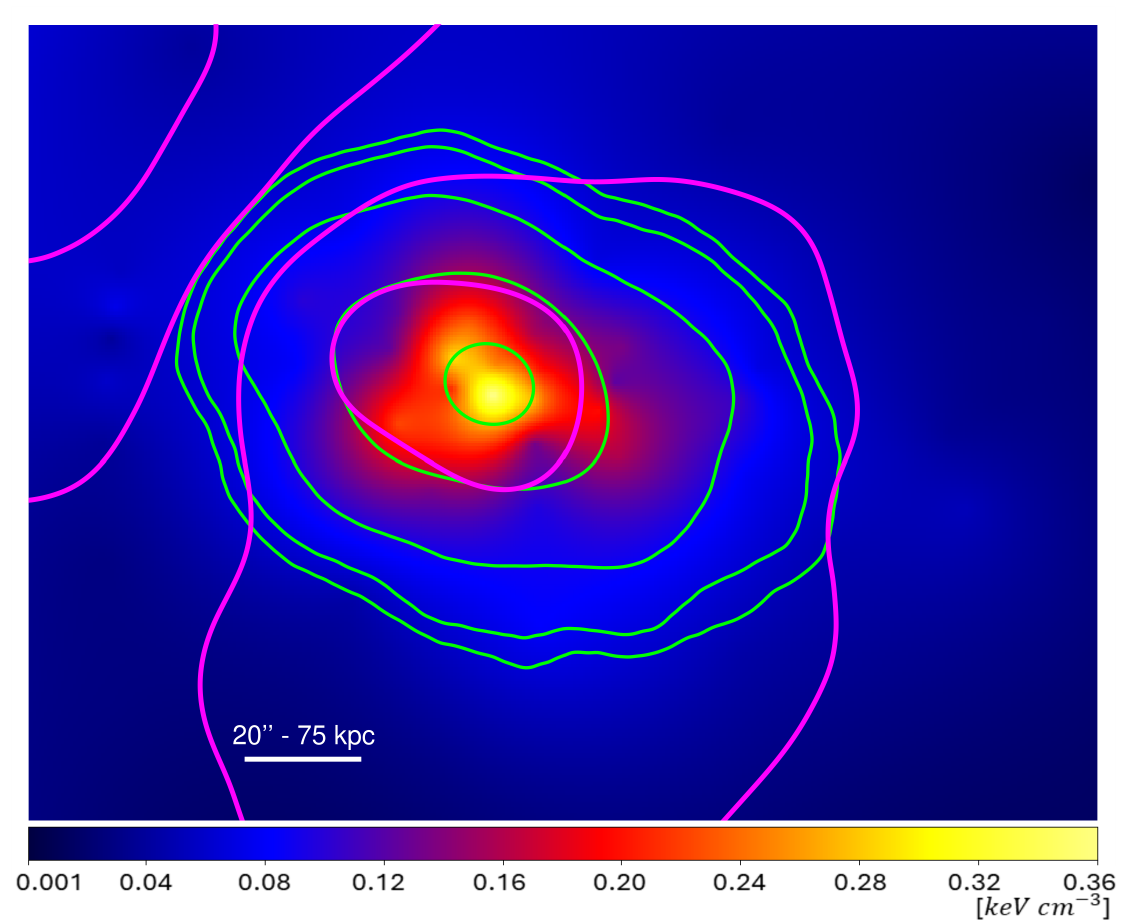}}
 \resizebox{6cm}{!}    {\includegraphics[width=\linewidth]{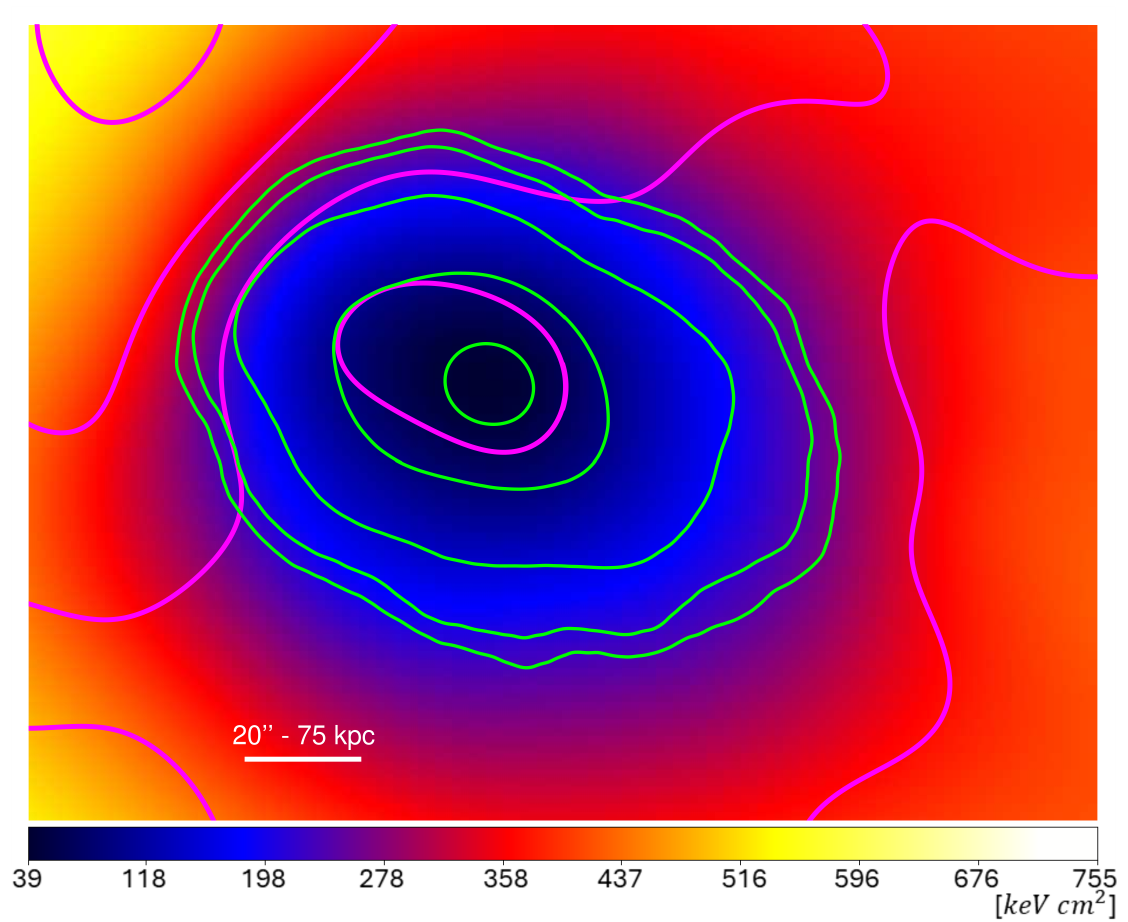}}
\caption{{Adaptive-smoothing maps of temperature (left), pressure (center), and entropy (right) for RXJ2129. In all three maps, the green contours represent the X-ray surface brightness isocontours and the magenta contours represent the relative errors for each parameter (see Sect. \ref{sect:x-rays}). In the temperature map, the contours delimit the boundaries inside which relative errors are 15\%, 20\%, 25\%, 30\%, and 35\% increasing from the center outwards. In the pressure map, contours represent 25\%, 30\%, 35\%, and 40\% relative error boundaries and, in the entropy map, 20\%, 27.5\%, 35\%, and 42.5\% relative error limits.}} \label{fig:adaptative_smooth_map}
\end{figure*}

Given the apparently minor misalignment between the cold front and the general X-ray/ICL elongation we produced an adaptive smoothed map of the pertinent thermodynamical quantities (Fig. \ref{fig:adaptative_smooth_map}). The adaptive smoothed map uses square cells, the size of which is determined by the minimal number counts contained within. {The projected temperature ($T$),  pressure ($P= n kT$), entropy ($S= kT n^{-2/3}$), and gas density $n$ were obtained from spectral fittings of the “cells”, the size of which is chosen in such a way as to have about 1000 counts each. The spectra of the cells were fitted using the same local background region mentioned earlier. The cells were distributed across the entire available cluster region. The results were smoothed using Kriging interpolation \citep[e.g., ][]{matheron_principles_1963}, which has been shown to highlight well relatively fine substructures \citep[e.g., ][]{dupke2007a,dupke2007b,jimenez-teja2023}. Relative errors for each one of these three maps are also shown in Fig. \ref{fig:adaptative_smooth_map} with green contours. The values of the relative errors within the contours vary from panel to panel. In the temperature map, the contours delimit the boundaries, inside which relative errors are 15\%, 20\%, 25\%, 30\%, and 35\%, increasing from the center outwards. For example, a 15\% relative error in the regions where the temperatures are about 6 keV would imply a value of $6.0 \pm 0.9$ keV. Analogously, the contours for the pressure map show the 25\%, 30\%, 35\%, and 40\% relative error boundaries and, for the entropy map, 20\%, 27.5\%, 35\%, and 42.5\%.} \\
 
\begin{table}
\centering
\begin{tabular}{lccc}
 Region & R$\pm\delta $R & T$_X\pm\delta $T$_x$ & $\chi^2/dof $\\
   & [kpc] & [keV] & [solar]\\ \hline
TF1 & 17$\pm$13 & 3.94$\pm$0.38 &  33.1/40\\
TF2 & 49$\pm$19 & 4.65$\pm$0.45 &  59.7/55\\
TF3 & 100$\pm$33 & 9.25$\pm$1.53 &  63/60\\
TF4 & 171$\pm$38 & 7.65$\pm$1.1 &  46.1/58\\
TF5 & 256$\pm$48 & 8.56$\pm$1.58 &  64.4/52\\
AF1 & -26$\pm$23 & 4.72$\pm$0.52 &  62.6/56\\
AF2 & -71$\pm$23 & 6.47$\pm$0.75 & 59/63\\
AF3 & -116$\pm$23 & 8.39$\pm$1.46 &  46.2/50\\
AF4 & -161$\pm$23 & 5.86$\pm$0.87 & 34.1/36\\
AF5 & -206$\pm$23 & 8.27$\pm$2.14 & 24.3/24\\
\end{tabular} 
\caption{Best-fit projected gas temperatures towards (positive) and away (negative) from the discontinuity} \label{bestfit_proj}
\end{table}

\section{Discussion} \label{sect:discussion}

\subsection{RXJ2129 dynamical state}\label{sect:discussion:dynamical_state}
RXJ2129 is one of 11 clusters that had their ICL analyzed in \cite{Jimenez-Teja2018}, using just HST optical images. {RXJ2129 ICL fractions (ratio of ICL to total light of the cluster) were low (ranging between 2 and 11\%) and nearly constant within the error bars, as those of relaxed clusters}. Contrarily, merging clusters had higher ICL fractions on average and a characteristic peak or excess in a certain wavelength range. Therefore, the ICL fractions suggested that RXJ2129 is a relaxed cluster that did not suffer any significant merging event in the recent past. This result confirmed the classification made by different morphological parameters measured over the hot gas distribution in X-rays \citep{donahue2016}, the presence of a cool core, or the identification of a mini-halo possibly linked to a strong radio source hosted by the BCG \citep{kale2015,pandey-pommier2016,giacintucci2017}. \\

Indeed, the X-ray analysis done here shows a surface brightness elongation similar to that of the optical data, including the ICL and shells with a position angle of $\sim 65^{\circ}$, consistent with that found by \cite{ueda}. The projected temperature profile confirms the presence of a cool core, where the temperature drops by a factor of $\sim 2$ from $\sim$ 8 keV at r~$\ge$ 100 kpc. The main surface brightness discontinuity to the west is $\sim 30^{\circ}$ from the main X-ICL aligned axies, and it is the most notable feature. It has the characteristics of a cold front, with no significant pressure discontinuity across the discontinuity. 
There are also mild indications of weaker cold fronts to the S-SE and N-NE directions as suggested by the image and gas pressure maps\footnote{A more detailed analysis of the gas core structures is beyond the analysis presented here and will be published elsewhere}. This suggests the presence of multiple sloshing cold fronts, as a consequence of previous nearby low mass galaxy system fly-bys or minor mergers,  frequently found in relaxed cool core clusters  \citep[e.g., ][and references therein]{markevitchrev,dupkea496,ascasibar,roediger}. This, combined with the very low entropy values in the center of the cluster, $\sim 16.0\pm2.4$ keV cm$^2$, strongly suggests that this system is a relaxed cool core cluster \citep{Ghizzardi}.\\

\subsection{Shells}\label{sect:discussion:shells}

The most widely accepted formation scenario of shells in galaxies advocates that these are formed by tidal debris of stripped galaxies, ejected by different mechanisms \citep{pop2017,pop2018}. As a consequence, shells would share a common origin with the ICL. Traditionally, the only mechanisms considered to explain the formation of shells were isolated minor mergers with a tight interval of infall orbits, with very low angular momentum \citep[i.e., radial infall trajectories; ][]{dupraz1986,herniquist1987,hernquist1988,hernquist1989,seguin1996,sanderson2010,bartoskova2011,ebrova2012}. However, the major merger scenario was also explored \citep{hernquist1992,gonzalez-garcia2005a,gonzalez-garcia2005b,gonzalez-garcia2005c,lima-neto2020}, and \cite{pop2018} showed that major mergers (defined as those with a mass ratio between the interacting galaxies greater than 1/10) are a more statistically probable formation pathway. Moreover, major mergers producing shells allow for a wider range of impact parameters because the efficiency of the dynamical friction is higher compared to minor mergers, although the most favourable cases are those were the radial component of the merger dominates the tangential one. However, the closer the mass ratio between the satellite and central galaxies, the less stringent the radial velocity condition. As for the infall time, shell-forming progenitors are mainly accreted between 4.2 and 7.6 Gyr before the shells are observed and were stripped between 1 and 4 Gyr before \citep{pop2018}. Lower accretion/stripping time antecedence does not have time to form shells, while larger values may have formed shells, but they would be phase-mixed by the time of the observations and, thus, unobservable.\\

Shells can be classified into type I, II, and III according to their morphology and alignment with the major axis of the BCG+ICL system \citep{wilkinson1987,prieur1989}. Type I shells are approximately aligned with this major axis, while type II are characterized for having wider opening angles and relatively random orientations in the projected space, and type III are undefined cases. In the case of RXJ2129, we clearly observe type I shells. 
Indeed, the morphology of the BCG+ICL system in RXJ2129 is strikingly similar to the simulated massive galaxy in Fig. 10 panel B of \cite{pop2018}, after a major merger with a high-mass ratio satellite ($\sim 6/10$) accreted  $\sim 6.6$ Gyr ago with a relatively radial orbit. According to this simulation, this satellite was stripped $\sim 2.6$ Gyr ago and produced shells, a wall, a clump, and secondary, more diffuse clouds of ICL at a lookback time of 0.98 Gyr that are extraordinarily similar to those observed in RXJ2129 (see Fig. \ref{fig:regions}). By this lookback time, the system would have a stellar mass of $\sim 2.2\times 10^{12}$, extremely similar to the mass derived from our SED fitting {(see Table \ref{table:SEDfitting_physicalproperties})}. It is thus tentative to speculate that RXJ2129 could have {reached} its current configuration via a similar recent merger tree.\\

\subsection{ICL stellar properties} \label{sec:stellar_properties}

Although CIGALE and Prospector fit different sets of data (CIGALE, only photometry; Prospector, {photometry} and spectroscopy), both codes yield consistent results. It is remarkable, though, that the inclusion of the MUSE spectra reduces significantly the uncertainties in the physical parameters derived (see Table \ref{table:SEDfitting_physicalproperties}, which reflects the importance and the need for using spectra in ICL studies.  An important, still open question in our SED fittings is that we do not know in advance whether there is dust in the ICL; that is why we left this parameter free. As recent simulations predict and it is supported by a few observational results, there are solid evidences of star formation in the ICL \citep[in situ ICL;][]{montenegro-taborda2023,ahvazi2024,barfety2022}, so it is reasonable to expect dust in the ICL. Unfortunately, the bluest region of the SED of RXJ2129 remains still unconstrained, so we cannot estimate the exact amount of extinction in our models. However, from Fig. \ref{fig:SEDfitting} we can infer that the impact in the overall fitting should be minimal.\\

The estimated stellar mass of the ICL in RXJ2129 is $log_{10}(M_*/M_{\odot})=11.81\pm 0.34$ and $12.33^{+0.04}_{-0.03}$, as yielded by CIGALE and Prospector, respectively, highly consistent with the $log_{10}(M_*/M_{\odot})\sim 11-12$ values found by \cite{morishita2017} for the six Hubble Frontier Field Clusters at $0.3<z<0.6$. RXJ2129 has a mass of $M_{500}=(8.07\pm 1.48)\times 10^{14}$ M$_{\odot}$, based on X-rays, and $M_{500}=(6.06\pm 1.33)\times 10^{14}$ M$_{\odot}$, based on the galaxy velocities \citep{logan2022}. Assuming a mass of $\sim 7.07\times 10^{14}$ M$_{\odot}$ and applying the stellar mass-halo mass relation derived by \cite{kravtsov2018} (calculated under the same \cite{Chabrier2003} IMF than our stellar masses, see Sect. \ref{sect:SEDfitting}), which is described by:
$$log_{10}\: M_{*,tot} = 0.59\times(log_{10}\: M_{500}-14.5)+12.71$$

we obtain a stellar mass of $M_{*,tot}=8.24\times 10^{12}$ $M_{\odot}$ for the BCG plus satellite galaxies. Thus, {the ICL stellar mass fraction is $\sim 7.8\pm 6.1$\% and $25.9^{+3.2}_{-2.8}$\%, according to CIGALE and Prospector results, respectively}. These values are in very nice agreement with the average ICL mass fraction of $10.9\pm 5$\% estimated from a sample of 683 clusters between $0.2<z<0.3$ by \cite{zibetti2005}. They are also consistent with the stellar mass fractions of 5\% to 20\% estimated by \cite{morishita2017} at higher redshift, which suggests that the ICL mass fraction does not evolve dramatically with redshift, at least in the interval $0.23<z<0.6$, in concordance with previous works \citep{krick2007}. Compared to previously measured optical ICL fractions (in flux) of RXJ2129 by \cite{Jimenez-Teja2018}, which range between 2 and 11\%, and considering that, at this redshift, the optical mostly traces young or low metallicity stars, which are not the main drivers of the total mass budget, our ICL mass fractions show a nice agreement. \\

The inferred ICL stellar mass fractions here are, however, lower than the ICL mass fractions predicted by \cite{contini2023} using semi-analytical models for highly-concentrated (thus, likely relaxed) clusters. However, given the uncertainties of the observational estimations of the virial mass, the stellar mass-halo mass relation, and the SED-derived ICL mass, along with the scatter in the simulations and the low statistics in the high-mass end, both observational and simulations results can be considered fairly consistent. Similarly, given the stellar/halo mass of RXJ2129, our ICL mass fractions would be located within the lower limit of the fractions estimated by \cite{contreras-santos2024}  (see their Fig. 5), \cite{proctor2024} (see their Fig. 11), \cite{yoo2024} (see their Fig. 4), and \cite{fu2024} (see their Fig. 10), using different simulations. \\

We found that the main bulk of ICL is composed by very old stars ($\sim 10$ Gyr), in concordance with several previous works. For example, \cite{melnick2012} calculated that the ICL of the cluster RXJ0054.0-2823, at $z=0.29$, was dominated by old stars, with $91\pm 3$\% of the mass being composed of $\sim 10$ Gyr stars. A compatible result was found in the Hydra I cluster (separated a time interval of 2.8 Gyr from RXJ2129) by
{\cite{coccato2011}}, whose ICL is mainly {composed} by $>13$ Gyr stars. A similar composition is found by \cite{edwards2016} in three {low-redshift} ($z<0.06$) clusters, with an ICL dominated in mass by stars of 12-13 Gyr and an increasing presence of younger stars as the degree of disturbance of the cluster was higher.\\

Nevertheless, we find a clear disagreement with the average value of $1-3$ Gyr found by \cite{morishita2017}, which cannot be explained by the time difference between their sample and RXJ2129 (their clusters are $\sim 2.3$ Gyr younger). All Frontier Field clusters are highly unrelaxed, so they had a relatively recent supply of young and/or low metallicity stars into the ICL from a past merger \citep{Jimenez-Teja2018,jimenez-teja2021,jimenez-teja2023,deoliveira2022,dupke2022}. Indeed, merging clusters show an ICL excess (IE) in the wavelength range that comprises the emission peaks A- and F-type stars, which are believed to be the responsible {for} this distinctive signature. Supporting this scenario,  \cite{morishita2017} found a meaningful amount of very young stars ($\sim 1$ Gyr) present in the ICL of the six clusters of their sample. In our case, the relaxed state of RXJ2129, confirmed independently by different indicators \citep[see Sect. \ref{sect:discussion:dynamical_state} and ][]{Jimenez-Teja2018}, would translate into an ancient ICL stellar population, that passively evolves since the last merging event (presumably, $\sim 6.6$ Gyr ago, with an infalling satellite galaxy that was stripped $\sim 2.6$ Gyr ago; see Sect. \ref{sect:discussion:shells}). Assuming an average age for an A/F-type star of $\sim 2.3$ Gyr using a Salpeter initial mass function from 1.1-2.5 $M_{\odot}$, and also assuming that they were extracted halfway through their lifetimes, these stars would have died since they were stripped. Only the extant stellar populations would still survive, contributing to the older age of the ICL in RXJ2129, compared to that of the Frontier Field clusters.\\


\section{Conclusions} \label{sect:conclusions}

We performed the first joint MUSE, HST, and JWST {analysis} of the ICL in a cluster, {to obtain} the most detailed ICL SED ever studied. The superb spatial resolution and depth of the optical and infrared imaging, combined with the optimal spectral coverage of the MUSE spectra in the optical, {allowed us to fit} the SED of the ICL in the cluster RXJ2129 from $\sim 0.4$ to 5 $\mu$m. Additionally, the incredible sensitivity of the HST+JWST images led us to unveil remarkable features in the ICL, such as shells, regions of lower density, and a {clump}, that tentatively allows us {elucidate} the recent past merger history of this cluster. Additionally, we analyzed the existing Chandra X-ray data to complement the dynamical history of the cluster from the study of the hot gas distribution. We summarize the main results:
\begin{itemize}
\item The X-ray analysis shows a cold front suggestive of a sloshing process after some past merger, confirming that RXJ2129 is currently in a relaxed state as previous works claimed using differen independent indicators \citep[remarkably, among them, the optical ICL fractions; ][]{Jimenez-Teja2018}.
\item We generated ICL maps for 15 HST and JWST images of RXJ2129, plus the 3681 frames of its MUSE-DEEP datacube. From this, we measured the total flux of the ICL, deriving its global physical properties via parametric and non-parametric SED fitting algorithms (CIGALE and Prospector). We estimate an ICL stellar mass $log_{10}(M_*/M_{\odot})=11.5-12.7$ and a main age of the older population between $9.7-10.5$ Gyr. These values are in agreement with  observational estimations calculated in the literature for other clusters. The only exception is a notable difference in age with respect to highly disturbed clusters, suggestive of different stellar populations that are direct consequence of different ICL formation channels.
\item The ICL mass fractions estimated from numerical simulations are in general higher than that observed in RXJ2129. However, given the uncertainties of the observational measurements and the large scatter and low statistics of some simulations in this mass regime, the results are fairly compatible.
\item Although a smooth distribution of ICL was expected for RXJ2129 given its relaxed stage, several features were discovered, such as the presence of several shells, clouds of ICL with different density, and a clump. The extraordinary resemblance of the BCG+ICL system in RXJ2129 with a simulation described in \cite{pop2018} led us to speculate a common formation mechanism: a major merger with a high-mass ratio satellite ($\sim 6/10$), accreted $\sim 6.6$ Gyr ago in a low angular momentum orbit and stripped $\sim 2.6$ Gyr ago.
\end{itemize}

These results show the potential of combining multiwavelength, deep, high quality data, to derive the global properties of the ICL and the dynamical stage of the host halo and the recent past accretion tree of galaxy clusters. Coupled with X-ray observations of the hot gas distribution, the potential of the ICL as tracer of the dynamical history of the cluster is enhanced. Given the superb quality of these data, the local properties and internal dynamics of the ICL and its different substructures will be studied in a forthcoming paper, opening the door to new applications as, for instance, understanding the role of the ICL in the chemical enrichment of the intracluster medium.\\

\begin{acknowledgements}

{We thank the anonymous referee for his/her kind, detailed, and helpful report, that has certainly improved the quality of this manuscript.} Y.J-T., A.G-A., and J.M.V. acknowledge financial support from the Spanish MINECO grant PID2022-136598NB-C32 and from the State Agency for Research of the Spanish MCIU through ‘Center of Excellence Severo Ochoa’ award to the Instituto de Astrofísica de Andalucía CEX2021-001131-S funded by MCIN/AEI/10.13039/501100011033; they also acknowledge financial support from the Spanish project PID2022-136598NB-C32. Y.J-T. acknowledges financial support from the European Union’s Horizon 2020 research and innovation programme under the Marie Skłodowska-Curie grant agreement No 898633 and the MSCA IF Extensions Program of the Spanish National Research Council (CSIC). R.A.D. acknowledges partial support from CNPq grant 312565/2022-4. P.K. acknowledges JWST GO-2767, and NSF AAST-2308051. This work was supported by research grants (VIL16599, VIL54489) from VILLUM FONDEN.

\end{acknowledgements}

\vspace{5mm}
\facilities{MUSE, HST(ACS), HST(WFC3), JWST(NIRCam)}



\bibliography{main.bbl}{}
\bibliographystyle{aasjournal}

\end{document}